\documentclass[fleqn,10pt]{wlscirep}
\usepackage[utf8]{inputenc}
\usepackage[T1]{fontenc}
\usepackage{lineno,hyperref}
\usepackage{graphicx}
\usepackage{multicol}
\usepackage{caption}
\usepackage{subcaption}
\usepackage{times}
\usepackage{tabulary,amsmath,graphicx,amsfonts,amssymb}
\usepackage{tabulary}
\usepackage[export]{adjustbox}
\usepackage{geometry}
\usepackage{hyperref}
\title{Deep learning framework for subject-independent emotion detection using wireless signals}
\author[1,+]{Ahsan Noor Khan}
\author[1,+]{Achintha Avin Ihalage}
\author[1,+]{Yihan Ma}
\author[1]{Baiyang Liu}
\author[1]{Yujie Liu}
\author[1,*]{Yang Hao}
\affil[1]{School of Electronic Engineering and Computer Science, Queen Mary University of London, E1 4NS, United Kingdom}
\affil[*]{y.hao@qmul.ac.uk}
\affil[+]{These authors contributed equally to this work}

\keywords{Keyword1, Keyword2, Keyword3}

\begin{abstract}
Emotion states recognition using wireless signals is an emerging area of research that has an impact on neuroscientific studies of human behaviour and well-being monitoring. Currently, standoff emotion detection is mostly reliant on the analysis of facial expressions and/or eye movements acquired from optical or video cameras. Meanwhile, although they have been widely accepted for recognizing human emotions from the multimodal data, machine learning approaches have been mostly restricted to subject dependent analyses which lack of generality. In this paper, we report an experimental study which collects heartbeat and breathing signals of 15 participants from radio frequency (RF) reflections off the body followed by novel noise filtering techniques. We propose a novel deep neural network (DNN) architecture based on the fusion of raw RF data and the processed RF signal for classifying and visualising various emotion states. The proposed model achieves high classification accuracy of 71.67\% for independent subjects with 0.71, 0.72 and 0.71 precision, recall and F1-score values respectively. We have compared our results with those obtained from five different classical ML algorithms and it is established that deep learning offers a superior performance even with limited amount of raw RF and post processed time-sequence data. The deep learning model has also been validated by comparing our results with those from ECG signals. Our results indicate that using wireless signals for stand-by emotion state detection is a better alternative to other technologies with high accuracy and have much wider applications in future studies of behavioural sciences.
\end{abstract}
\begin{document}

\flushbottom
\maketitle
%
%
\thispagestyle{empty}

\section*{Introduction}
With the advancements in body-centric wireless systems, physiological monitoring has been revolutionized for improving healthcare and wellbeing of people \cite{10.5555/1204228,6863652,8329985,5738699,WANG2014406,dias2018wearable}. These systems predominantly rely on wireless intelligent sensors that are capable of retrieving clinical information from physiological signals to interpret the progression of various ailments. A traditional sensing system can amass and process a wide range of biological signals, including electrophysiological (electroencephalogram (EEG), electrocardiogram (ECG)) \cite{doi:10.1002/adma.201301921,s101210837} and physiological information \cite{schwartz2013flexible,webb2013ultrathin}, that are incessantly emanating from the human body.  Apart from diagnostic and therapeutic aspects, wireless sensors have demonstrated their applications for recognizing emotions that can be extracted from a measured physiological data \cite{zhao2016emotion,8105799,5738690}. Emotions are indispensable facet of humans and can affect their physiological status during office work, travelling, decision making, entertainment and many others activities \cite{Dolan1191,shu2018review}. The human health and work productivity are highly reliant on the intensity of emotions that can be either positive or negative. The positive emotions can help to achieve optimal well being and mental strength, whereas long term negative emotions may result in predisposing cause of chronic mental health problems, such as depression and anxiety. Furthermore, people who are experiencing frequent negative emotional states have a weaker immune response as compared to people with positive affective style \cite{rosenkranz2003affective}. The above mentioned aspects of emotions have led to further investigations in real life scenarios. 
\begin{figure*}[t!]
		\centering 
		\includegraphics[width=13cm]{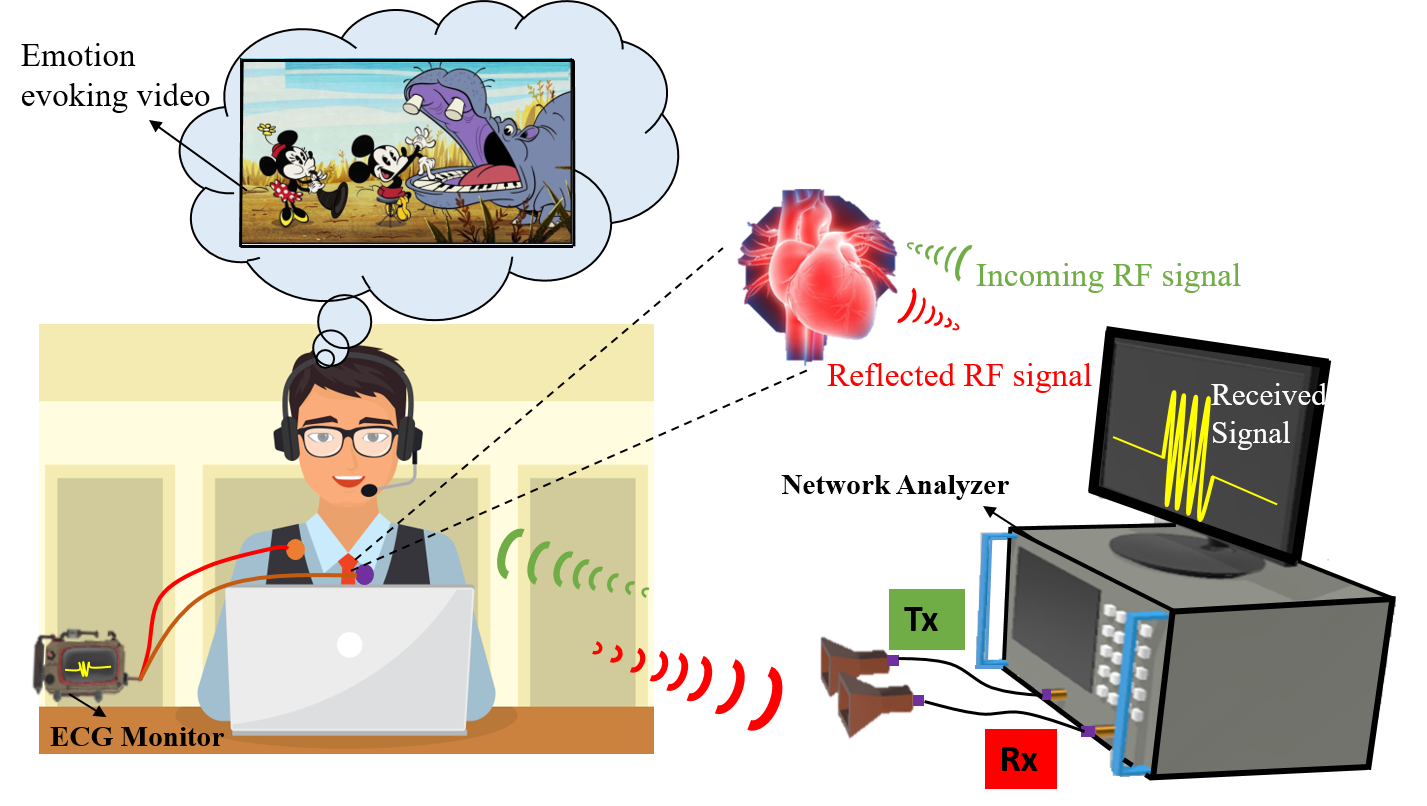} 
		\caption[caption]{Emotion detection process in which each participant is asked to watch emotion evoking videos on the monitor while being exposed with radio waves. The Tx antenna is used to transmit RF signals towards the participant, whereas Rx antenna is used to receive RF reflections off the body. The ECG monitor is also connected to a participant's chest for recording heart beats. The data received from ECG is used to correlate heart beats variations with emotion evoking videos.}
		\label{Dummy}
	\end{figure*}
\\ \noindent
In recent past, emotion recognition has been considered as a key research area in the fields of social security \cite{verschuere2006psychopathy}, multimedia entertainment \cite{mandryk2006using}, vehicle transport \cite{7831367}, brain computer interface \cite{lopez2019towards} and healthcare \cite{Rudoviceaao6760,7536936,8105799}. For instance, people experiences and emotions while watching a film in cinema can benefit film makers and producers to ameliorate film attributes, such as video quality, impact, social awareness, story line and actors performances. The tourism industry can employ sensors in the travelling bus to analyze visitors emotions while they are watching tourist attractions \cite{kim2017measuring}. The retrieved data from sensors can be conducive to explore lucrative destinations as tourists emotions are highly dependent on the places they visit. The marketing companies can estimate the impact of advertisements on emotions to increase their product sales and awareness. The healthcare professionals can understand the impact of therapy and diagnosis on the behaviour of patients. Furthermore, there have been considerable efforts to detect emotions through brain-computer interfaces \cite{kashihara2014brain,7727453}. \\
Due to the impact of aforementioned applications in our daily course of life, extensive amount of strategies have been exploited for emotion detection that primarily focus on audio \cite{4468714}, visual \cite{Krageleaaw4358,4468714,kahou2016emonets}, facial \cite{TARNOWSKI20171175}, speech \cite{ELAYADI2011572} and body gestures \cite{8887272}. The recent progress in wearable electronic sensors have enabled collection of physiological data, such as heart rate, respiration rate and electroencephalography (EEG ) for several physical manifestations of emotions. However, wearable sensors and devices are cumbersome during routine activities and can lead to false judgement in recognizing people's true emotions. In \cite{zhao2016emotion}, a wireless system is demonstrated that can measure minute variations of a person's heartbeat and breathing rate in response to the individually prepared stimuli (memories, photos, music and videos) that evoke a certain emotion during experiment. Most of the participants in the study were actors and experienced in evoking emotions. The RF reflections off the body are preprocessed and fed to machine learning (ML) algorithms to classify four basic emotions types, such as anger, sadness, joy and pleasure. The proposed system excludes the requirements of carrying on-body sensors for emotion detection. Nevertheless, emotions were classified only using conventional ML algorithms and the quest to investigate the competence of deep learning for wireless signals classification has become an exciting research area. 
\\ \noindent
This paper focuses on exploring deep neural networks for affective emotion detection in comparison to traditional ML algorithms. A framework is developed for recognizing human emotions using a wireless system without bulky wearable sensors, making it truly non-intrusive, and directly applicable in future smart home/building environments. An experimental database containing the heartbeat and breathing signals of 15 subjects was created by extracting the radio frequency (RF) reflections off the body followed by noise filtering techniques. The RF based emotion sensing systems (Fig.~\ref{Dummy}) can overcome the limitations of traditional body worn devices that can encounter limited range of sensing and also cause inconvenience to people. For eliciting particular emotion in the participant, four videos have been selected from an on-line video platform. Videos were not shown to the participants before the start of experiment. Thus, our approach of evoking emotions in the participants is distinguishable from \cite{zhao2016emotion}, in which each participant has to prepare their own stimuli (after watching photos, reminding personal memories, music, videos) before the start of the experiment and act the intended emotion during the experiment. A novel convolutional neural network (CNN) architecture integrated with long short-term memory (LSTM) sequence learning cells that leverage's both the processed RF signal and raw RF reflection is utilized for the classification.
\begin{figure*}[t!]
		\centering 
		\includegraphics[width=16cm]{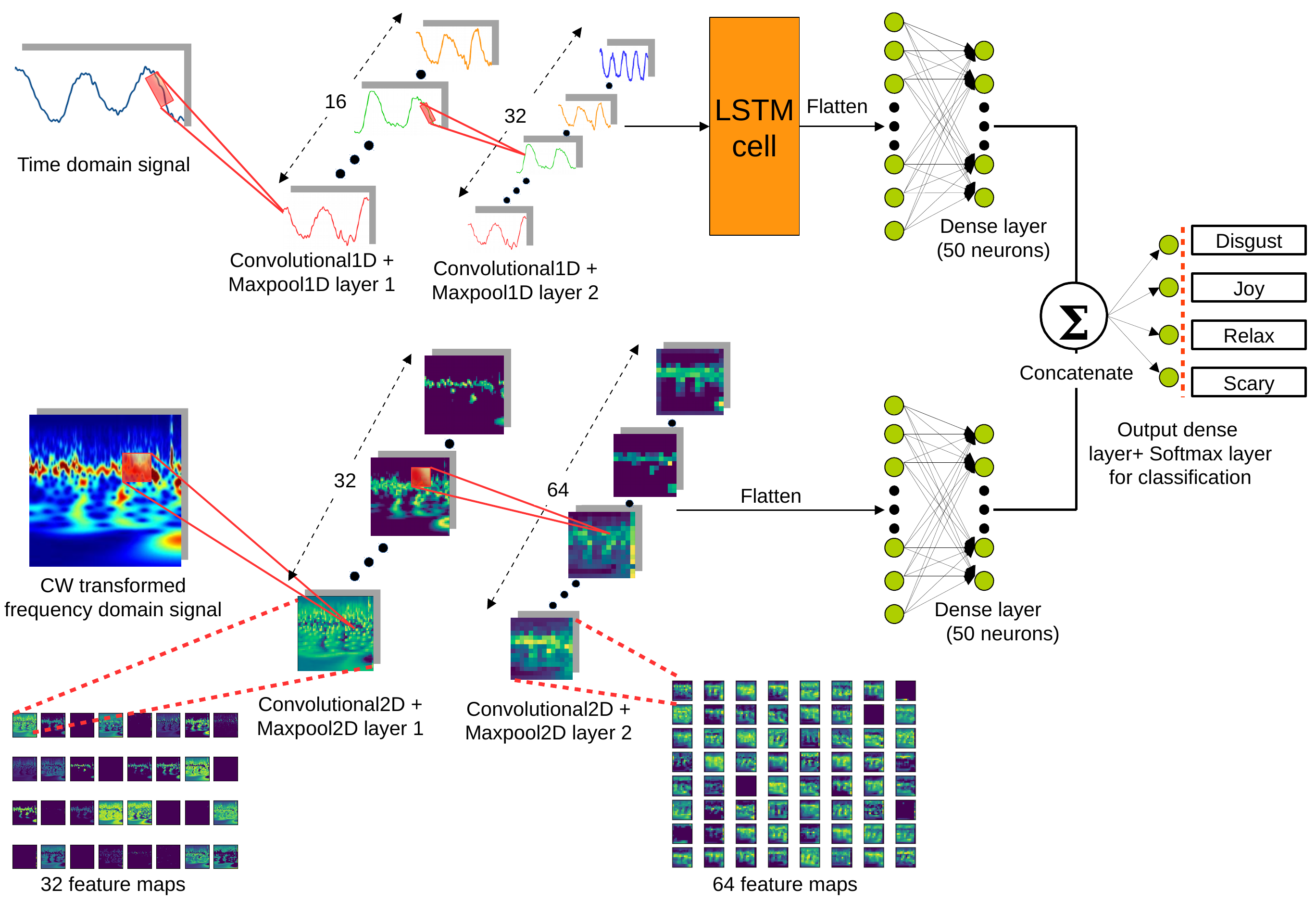} 
		\caption[caption]{Proposed deep neural network architecture for emotion classification. Time domain RF signal is processed through two convolutional-1D layers and an additional LSTM cell that captures the time dependency (supplementary information, section 1). The CW transformation is processed by two convolutional-2D layers (supplementary information, section 2.1). Each feature map in convolutional layers represents a unique extracted feature from the layer input. The features extracted from two distinct inputs of the model are then concatenated, leading to a broad learning capability. The detailed visualization of 32 and 64 features maps is presented in section 2.2 of supplementary information.}
		\label{DL_arch}
	\end{figure*}
\\ \noindent
The proposed network achieves state-of-the-art classification accuracy in comparison to five different traditional ML algorithms. On the other hand, a similar architecture is used for emotion recognition using the ECG signals. Our results indicate that deep learning is capable of utilizing a range of building blocks to learn from the RF reflections off the body for precise emotion detection and excludes manual feature extraction techniques. Furthermore, we propose that RF reflections can be an exceptional alternative to ECG or bulky wearables for subject-independent human emotion detection with high and comparable accuracy.
\section*{Results}

\subsection*{Detection of Emotional States}
\subsubsection*{Deep Learning Analysis}
\begin{figure*}[b!]
		\centering
		\begin{subfigure}{0.33\textwidth}
			\centering
			\includegraphics[width=1\linewidth]{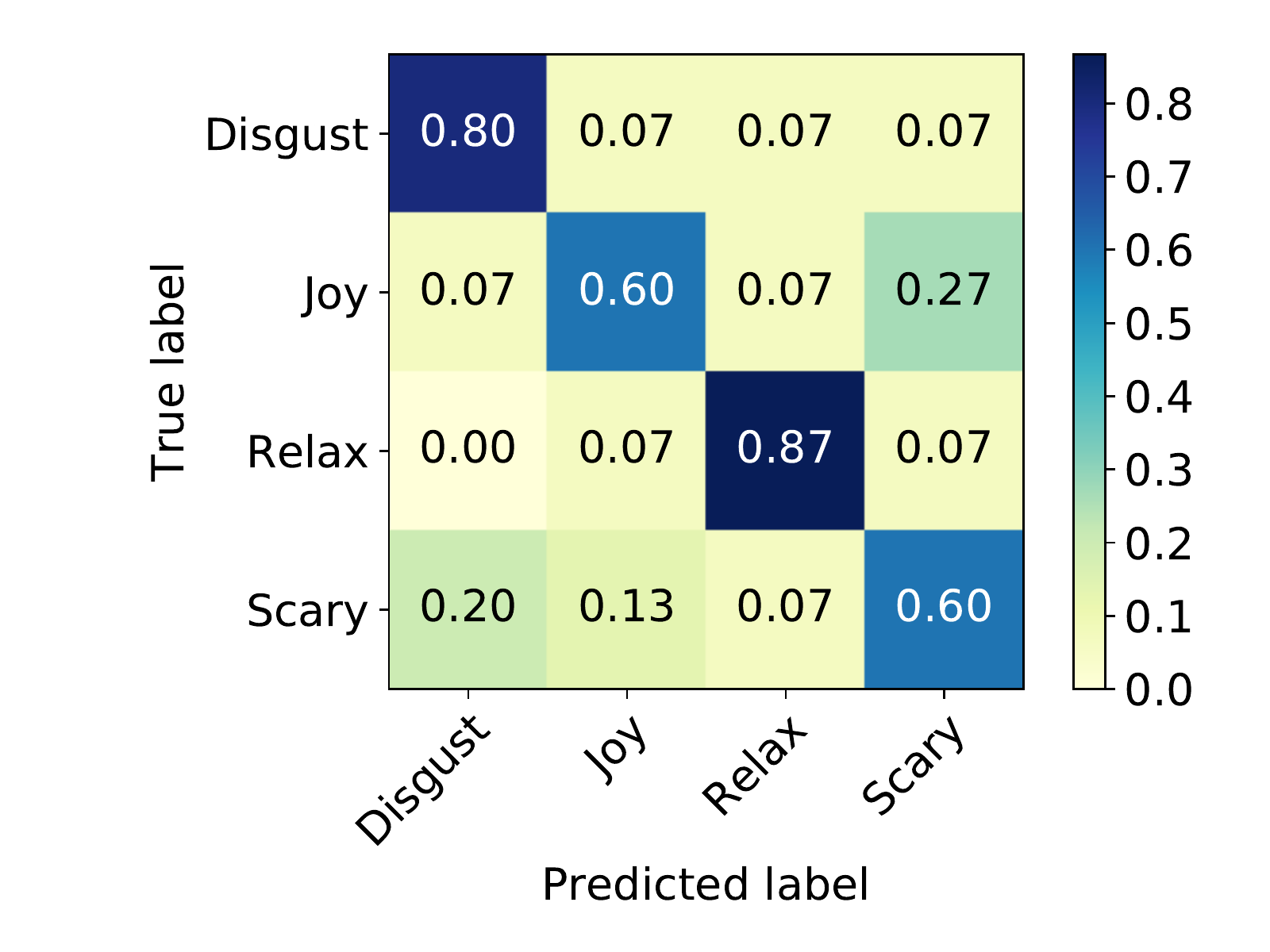}
			\caption{CNN + LSTM}
			\label{dl_conf}
		\end{subfigure}%
		\begin{subfigure}{0.33\textwidth}
			\centering
			\includegraphics[width=1\linewidth]{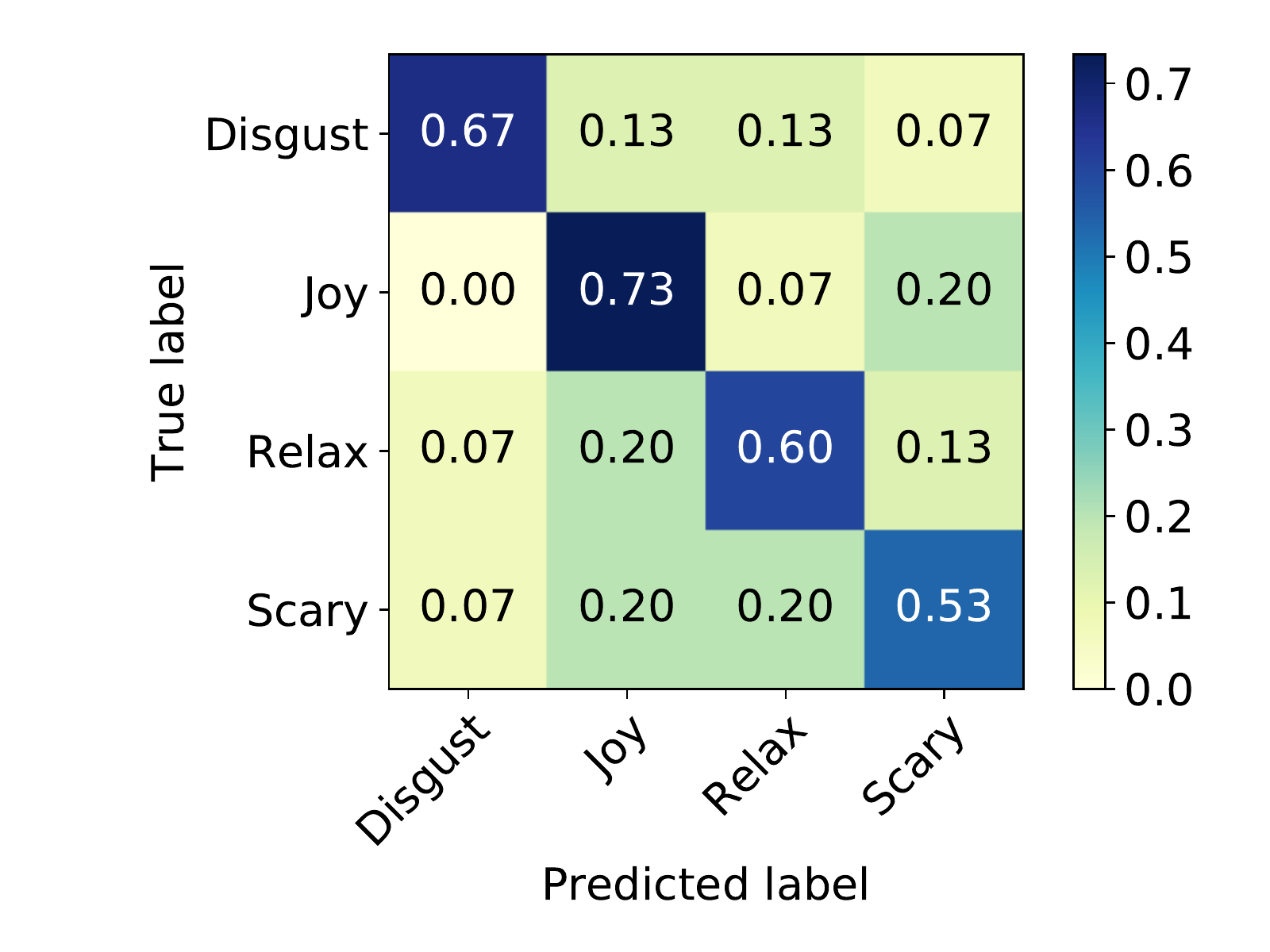}
			\caption{Random forest}
			\label{rf_conf}
		\end{subfigure}%
		\begin{subfigure}{0.33\textwidth}
			\centering
			\includegraphics[width=1\linewidth]{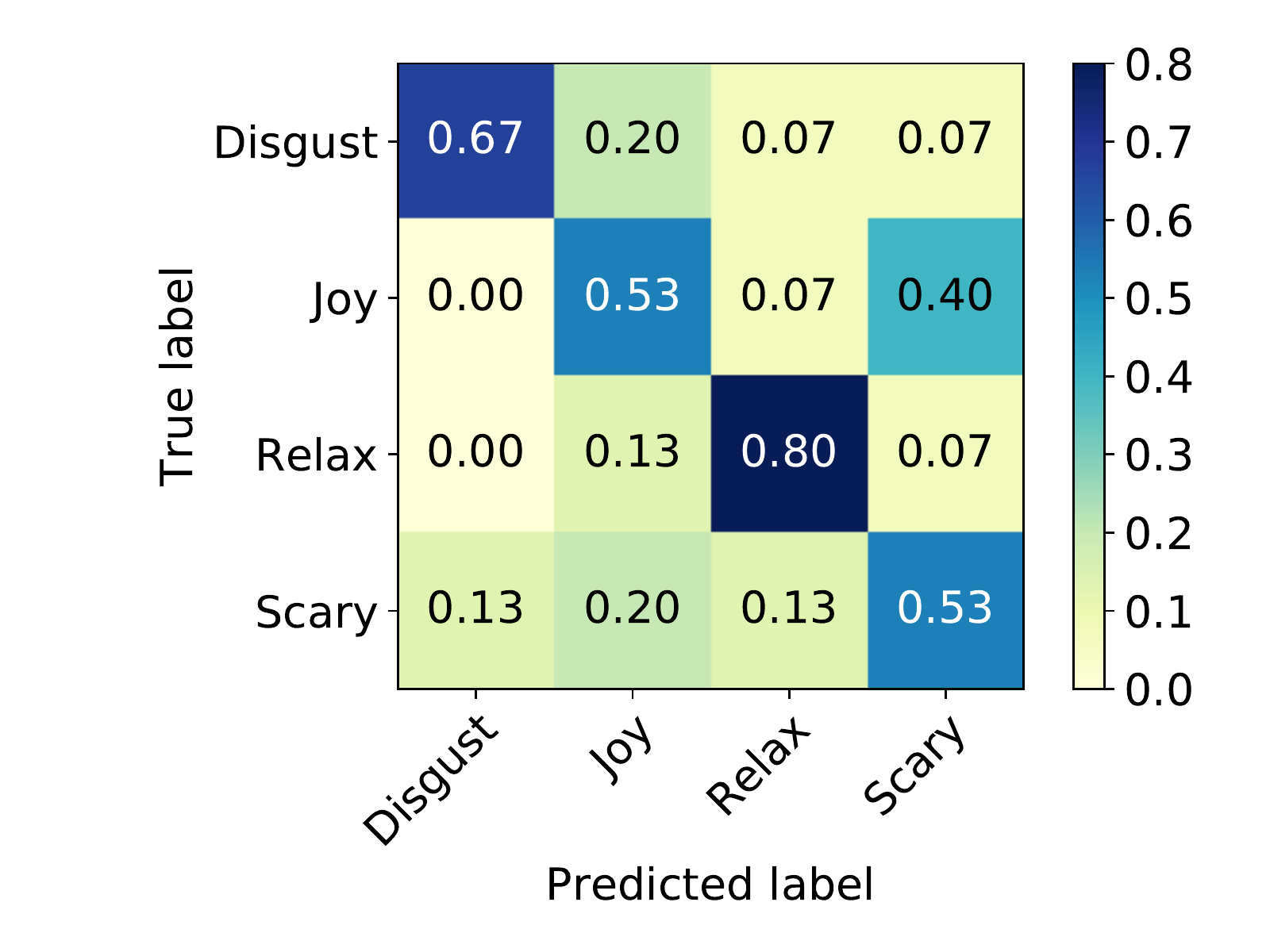}
			\caption{SVM}
			\label{svm_conf}
		\end{subfigure}%
		\vskip3ex
		\begin{subfigure}{0.33\textwidth}
			\centering
			\includegraphics[width=1\linewidth]{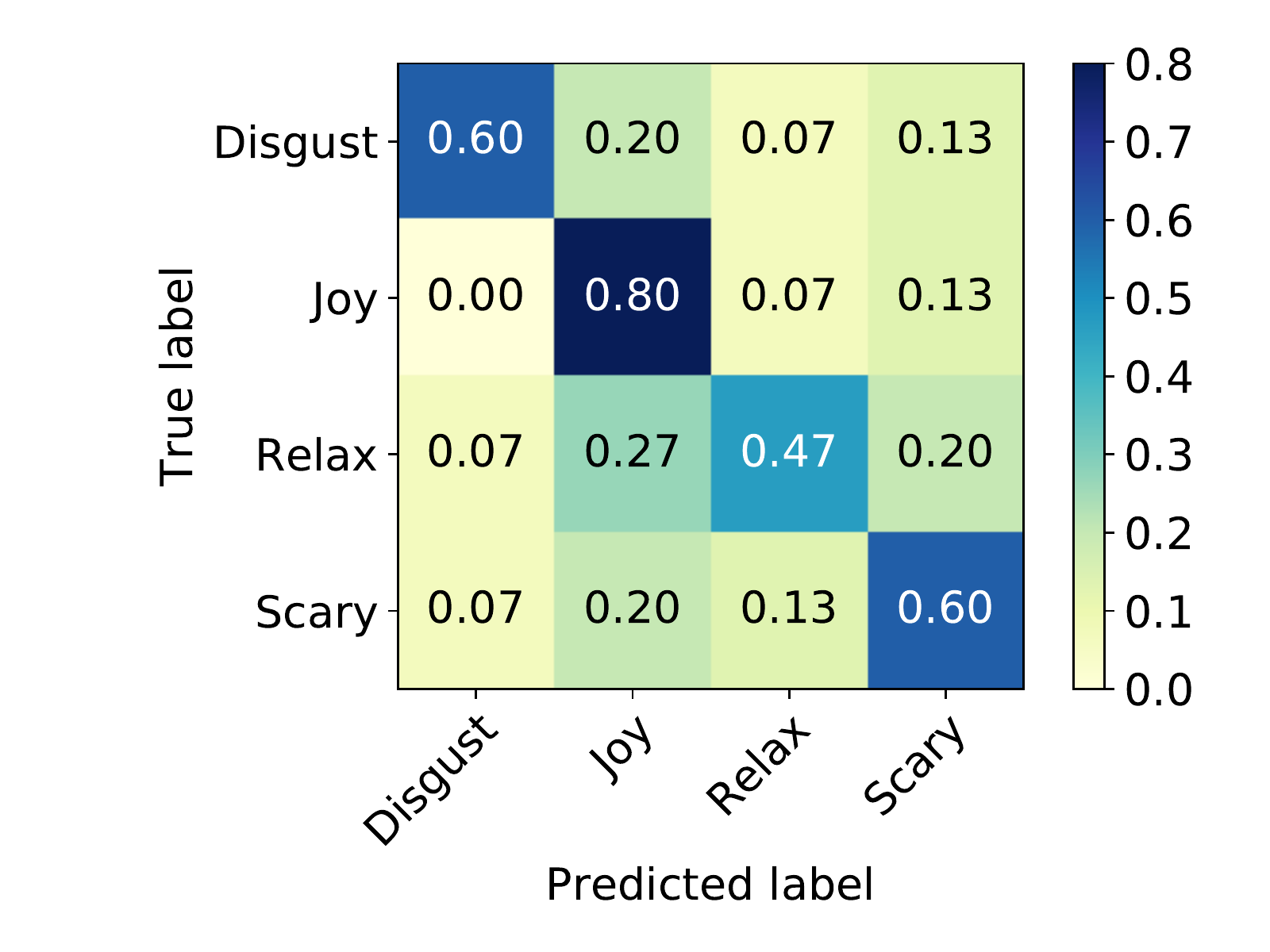}
			\caption{KNN}
			\label{knn_conf}
		\end{subfigure}
		\begin{subfigure}{0.33\textwidth}
			\centering
			\includegraphics[width=1\linewidth]{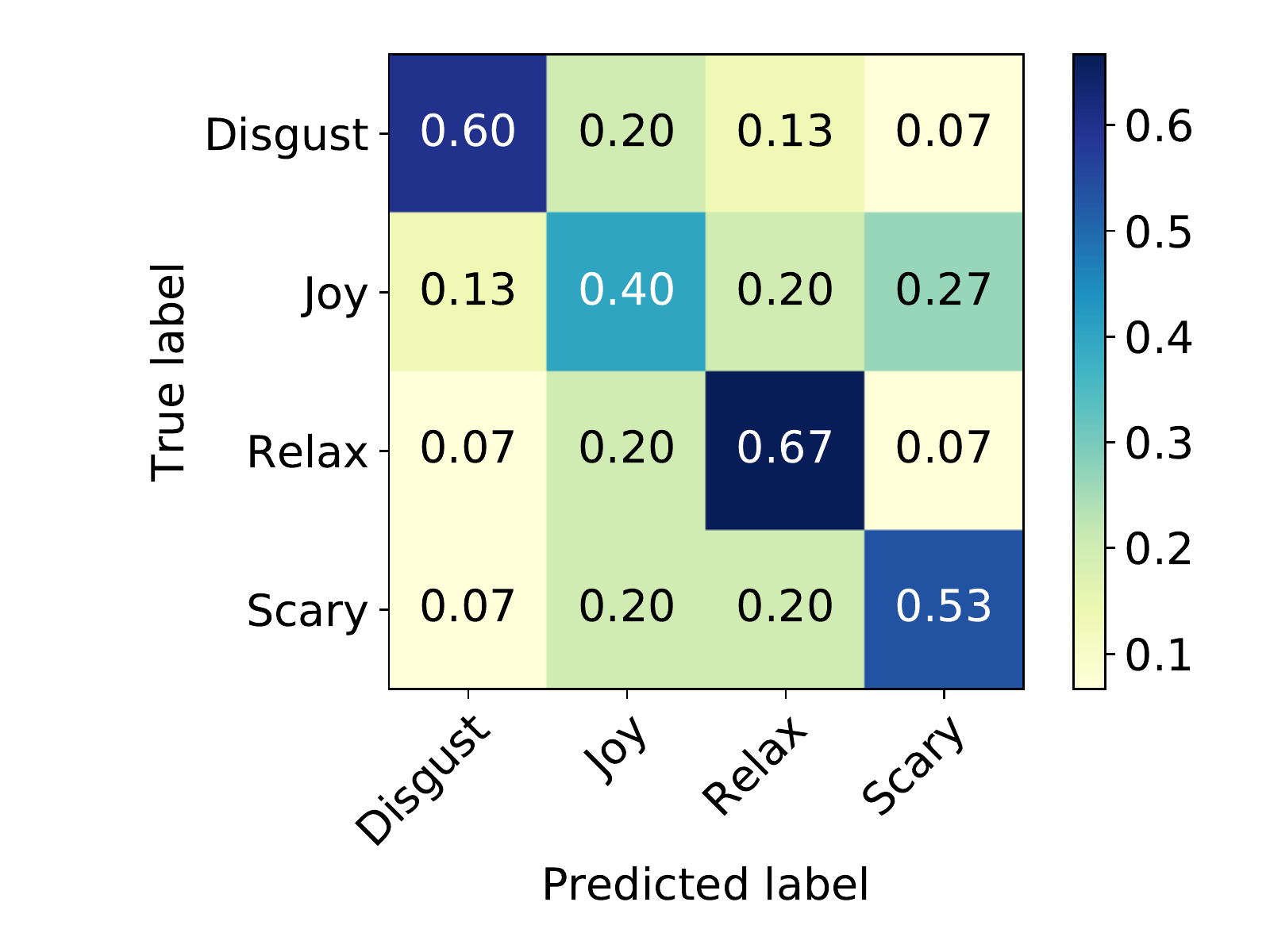}
			\caption{Decision tree}
			\label{dt_conf}
		\end{subfigure}%
		\begin{subfigure}{0.33\textwidth}
			\centering
			\includegraphics[width=1\linewidth]{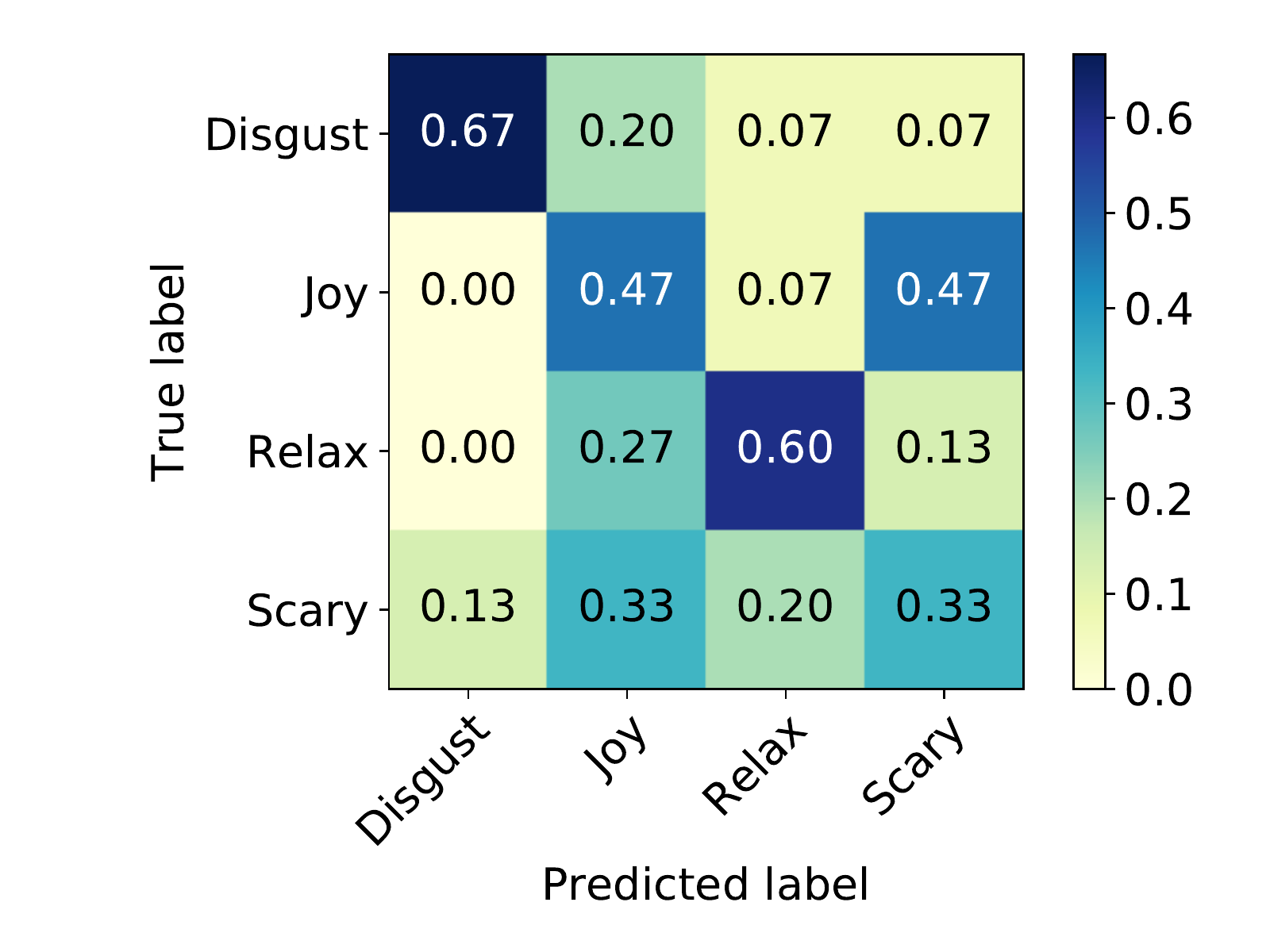}
			\caption{LDA}
			\label{lda_conf}
		\end{subfigure}
		\caption{Confusion matrices obtained by LOOCV for DL and ML models.}
		\label{conf_mats}
	\end{figure*}
Feature extraction is an integral part of a signal (electromagntic, acoustic, etc.) classification that can be performed manually or by using a neural network. Deploying traditional machine learning algorithms for signal classification necessitates ponderous extraction of statistical parameters from the raw data input. However, this manual approach can be tedious and may result in omission of some useful features. In contrary, deep neural networks can extract enormous amount of features from the raw data itself, whether they are significant or of minute details \cite{6889383}. Therefore, we employ an appropriate DNN architecture to process the time domain wireless signal (RF reflections off the body) and the corresponding frequency domain version obtained by continuous wavelet (CW) transformation. Here, the RF reflection signal is one-dimensional (1D) and the CW transformation is an image of three dimensions (3D), represented in the format of (width, height and channels). The parameters in wavelet image can be regarded as time (x-axis), frequency(y-axis), and the amplitude. The proposed DL architecture that is shown in Fig.~\ref{DL_arch} could be identified as a `Y' shaped neural network that accepts inputs in two distinct forms and fuses the processed inputs at the end to produce classification probabilities related to four emotions. The neural network consists of two sets of convolutional 1D and  maxpooling 1D layers, followed by a long short-term memory (LSTM) cell to capture the features and time dependency of the time domain RF signal. Another two sets of convolutional 2D and maxpooling 2D layers are used to process CW transformed image.
\\ \noindent
The convolutional layers are exceptional feature extractors and often outperform humans in this regard. A convolutional layer may have many kernels in the form of matrices (e.g. 3 $\times$ 3 and 5 $\times$ 5) that embed numerical values to capture variety of different features (e.g. brightness, darkness, blurring, edges, etc., of an image) from raw data. A kernel runs through the input data as a sliding window, and at every distinct location, it performs element-wise multiplication with the overlapping input data and takes the summation to obtain the value of that particular location of the generated feature map. Maxpooling layers do not involve in feature extraction. However, they reduce the dimensions of the outputs of convolutional layers, hence reducing the computational complexity. A typical convolutional layer has 32, 64 or even 128 kernels and thus results in the same number of feature maps. As observed in Fig.~\ref{DL_arch}, the feature maps carry even the diminutive information available in the input image, whereas a human eye is unable to capture this level of information, making them ordinary feature extractors.
\\ \noindent The accuracy of classification is evaluated with leave-one-out cross validation (LOOCV) \cite{kohavi1995study}. Although, cross validation is immensely used with ML models to observe the generalizability of the model, it is somewhat unconventional to perform cross validation with deep learning due to; (1) extreme computational complexity and (2) difficulty in tracking overfitting/underfitting conditions with a fixed number of iterations while training the model. However, in order to make a fair judgement on our DL predictions, we first used the full database and performed LOOCV, despite being the most computational intensive form of K-fold cross validation. In K-fold cross validation, the database is split into $k$ subsets, out of which, one is kept as the test set and the other $k-1$ are put together to form the training set. This process is repeated $k$ times such that every data point gets to be in the test set exactly once, eliminating the effect of biased data division into train and test sets. LOOCV is achieved by making the value of $k$ equal to $N$, number of data points in the database. 
\\ \noindent The proposed DL model yielded in 71.67\% LOOCV accuracy. This is quite a high percentage, considering the fact that human emotions are highly dependent on the level of stimulation generated in their brains by the same audio-video stimuli, capable of inducing emotions intensity differently from one person to another. It is tempting to conclude that the performance of model is solely based on the classification accuracy. However, a model with a high classification accuracy can still perform suboptimally, especially when the database is unbalanced as some classes contain a high number of data points and the others do not. In order to have a better description of the model, we often adopt other performance metrics such as precision, recall and F1-score. Precision indicates how many selected instances are relevant (a measure of quality), whereas recall indicates how many relevant instances are selected (a measure of quantity). F1-score reveals the trade-off between precision and recall, and can be correlated with effective resistance of the two parallel resistors (precision and recall) in a closed loop circuit. F1-score becomes low if either of these figures is low in comparison to the other, thus illustrating the reliability of the model across all classes. Although, these parameters are defined for binary classification, they can be extended to multi-class problems by calculating inter-class mean and standard deviation. The calculated values of precision, recall and F1-score after LOOCV are 0.713, 0.716 and 0.714 respectively, implying that the model has achieved good generalizability.
\subsubsection*{Machine Learning (ML) Analysis}
We have employed traditional ML algorithms process by means of data pre-processing, feature extraction, model training and classifications (supplementary information, section 3, Fig. S5). In our experiment, the RF reflected signals off the body encompass human body movements and random noise that is mostly contributed from the environment, equipments (VNA, cables, etc,...) and other moving objects. For this reason, it is essential to filter the noise from received RF signals for further processing. Moreover, we have also implemented data normalization technique to circumvent the influence of intensity variations on body movement for each participant. \\
Feature extraction process can be regarded as a core step of ML algorithms to analyse data. Considering the importance of ML for feature extraction, an efficient algorithm can significantly improve the classification accuracy while reducing the impact of interfering redundant RF signals and random noise. In the literature, a variety of feature extraction parameters are studied that are mostly in the field of affective recognition and biological engineering \cite{7887697, 6320605, SABETI2009263}. Permutation entropy is a widely used nonlinear parameter to evaluate the complexity of sequence that is a prevalent approach to estimate the pattern of biological signals, such as Electrocardiogram (ECG) and electroencephalogram (EEG). It is also capable of detecting real-time dynamic characteristics, and also has strong robustness.\\
Apart from the entropy value, it is well documented that the power spectral density (PSD) and statistical (variance, skewness,kurtosis) parameters are also related to the affective state of participants \cite{davidson2003affective}. In our analysis, the permutation entropy, PSD in the range of 0.15 - 2 Hz, 2 - 4 Hz and 4 - 8 Hz, and the variance, skewness and kurtosis values are extracted from the pre-processed signals. Therefore, overall seven parameters are tapped in the feature extraction process (supplementary information, section 3, Fig. S5). 
\subsubsection*{Analysis of Deep Learning and Machine Learning Results}
\begin{table*}[t!]
		\caption{{ML vs DL results comparison based on average performance metrics. The metric 'Accuracy' refers to LOOCV accuracy.} }
		\label{perf_metrics}
		\def\arraystretch{1}
		\ignorespaces 
		\centering 
		\begin{tabulary}{\textwidth}{p{\dimexpr.20\linewidth-2\tabcolsep}p{\dimexpr.20\linewidth-2\tabcolsep}p{\dimexpr.20\linewidth-2\tabcolsep}p{\dimexpr.20\linewidth-2\tabcolsep}p{\dimexpr.20\linewidth-2\tabcolsep}}
			\hline 
			   &  Accuracy (\%) &  Precision &  Recall  & F1-score\\
			\hline 
			CNN + LSTM &
			71.67 &
			0.713 ($\pm0.08$)&
			0.716 ($\pm0.12$)&
			0.714 ($\pm0.10$)\\
			Random forest &
			63.33 &
			0.646 ($\pm0.27$)&
			0.633 ($\pm0.29$)&
			0.634 ($\pm0.18$)\\
			SVM &
			63.33 &
			0.645 ($\pm0.17$)&
			0.63 ($\pm0.04$)&
			0.637 ($\pm0.08$)\\
			KNN &
			61.7 &
			0.64 ($\pm0.21$)&
			0.616 ($\pm0.18$)&
			0.615 ($\pm0.19$)\\
			Decision tree &
			55.0 &
			0.554 ($\pm0.30$)&
			0.549 ($\pm0.23$)&
			0.55 ($\pm0.14$)\\
			LDA &
			51.7 &
			0.544 ($\pm0.36$)&
			0.516 ($\pm0.27$)&
			0.526 ($\pm0.28$)\\
			\hline 
		\end{tabulary}\par 
		\label{comp_DLML}
	\end{table*}
The confusion matrices obtained using LOOCV for CNN+LSTM model and five classical ML algorithms are depicted in Fig.~\ref{conf_mats}. As tabulated in Table \ref{perf_metrics}, deep learning outperforms conventional machine learning algorithms in all performance metrics. We identify two main reasons that explain why deep learning is superior in the current learning problem. First, having both the time domain wireless signal and CW transformed image as an input is a rich source of learning for the CNN+LSTM model whereas the ML algorithms are trained with extracted features as inputs, that are sensitive to the level of human judgement on selecting features as well as the obvious loss of information from the original data. Second, CNNs are self learners that learn even the diminutive information, hidden in raw data that aids to reconstruct its target values, given the correct hyper-parameters. ML algorithms are somewhat reliant on human to figure out meaningful statistical parameters (or a combination of parameters) from raw data to be fed to the model. Nevertheless, these ML models still report an acceptable performance that can be used as a criterion for measuring how well the implemented DL model can perform. 
\subsection*{Data Visualization}
\begin{figure}[b!]
		\centering
		\begin{subfigure}{0.5\columnwidth}
			\centering
			\includegraphics[width=1\linewidth]{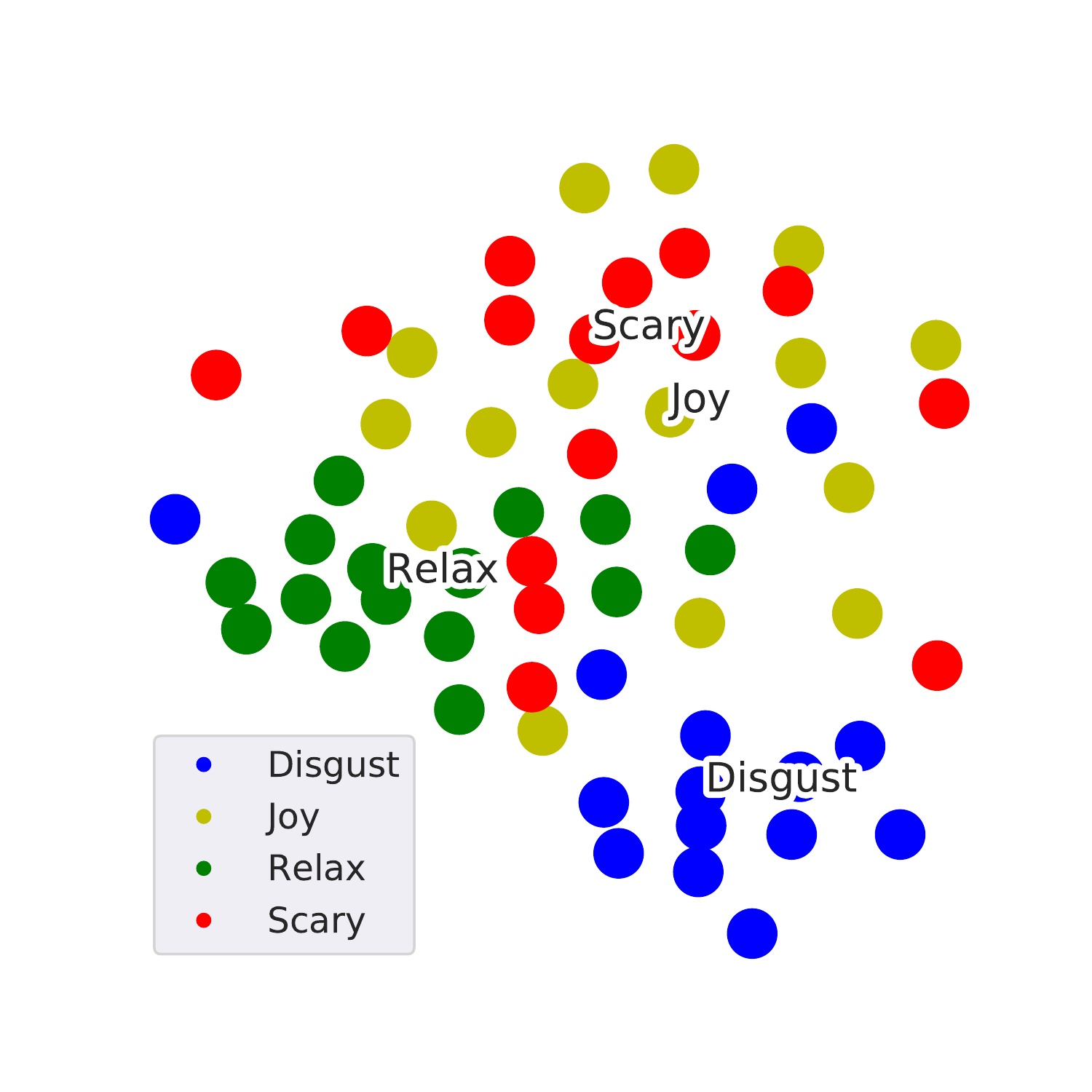}
			\caption{t-SNE plot of the RF data}
			\label{tsne_plot}
		\end{subfigure}
		\begin{subfigure}{0.5\columnwidth}
			\centering
			\includegraphics[width=1\linewidth]{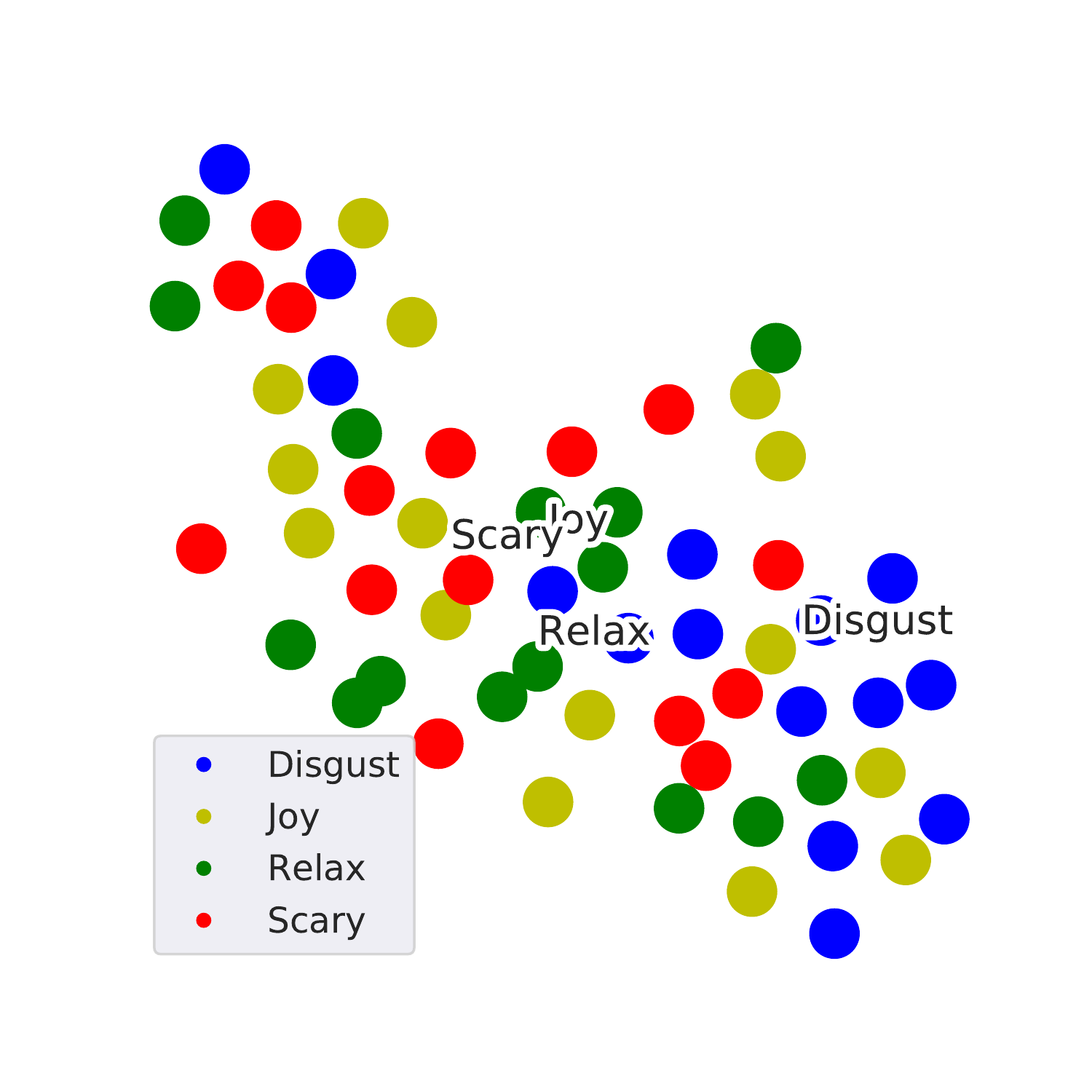}
			\caption{t-SNE plot of the ECG data}
			\label{ecg_tsne_plot}
		\end{subfigure}%
		\caption{t-SNE plots representing the full RF and ECG databases. The plots were obtained by reducing the dimensions of the continuous wavelet images of each signal. It can be observed that the wavelet images of RF signals (panel (a)) demonstrate a better separability between emotions than that of ECG signals (panel (b)). }
		\label{tsne_plots}
	\end{figure}	
Data visualization is pivotal for basic identification of patterns and trends in data that helps to understand and elaborate the results obtained from the machine learning models. However, high dimensional data as obtained by feature extraction, needs to be compressed into a lower dimension for visualization. T-distributed stochastic neighbour embedding (t-SNE) is a nonlinear dimensionality reduction machine learning algorithm often used for visualising high dimensional data by projecting it onto a 2D or 3D space (supplementary information, section 4). Fig.~\ref{tsne_plots} shows the t-SNE plots of RF and ECG databases, representing all 15 subjects. Fig.~\ref{tsne_plot} supports ML classification results for RF signals as the emotions `Relax' and `Disgust' are rather easily separable from the rest of emotions. The self assessment scaling acquired from the subjects after participating in the experiment also complement our findings as most of them stated that the video stimuli of `Relax' and `Disgust' really evoked the calmness and disgust emotions respectively (supplementary information, section 5, Fig. S6). 
\section*{Discussion}
It is understood that the emotions evoked by the audio-visual stimuli are highly subject dependent and therefore difficult to classify on a common ground.
\begin{figure*}[h!]
		\centering 
		\includegraphics[width=12cm]{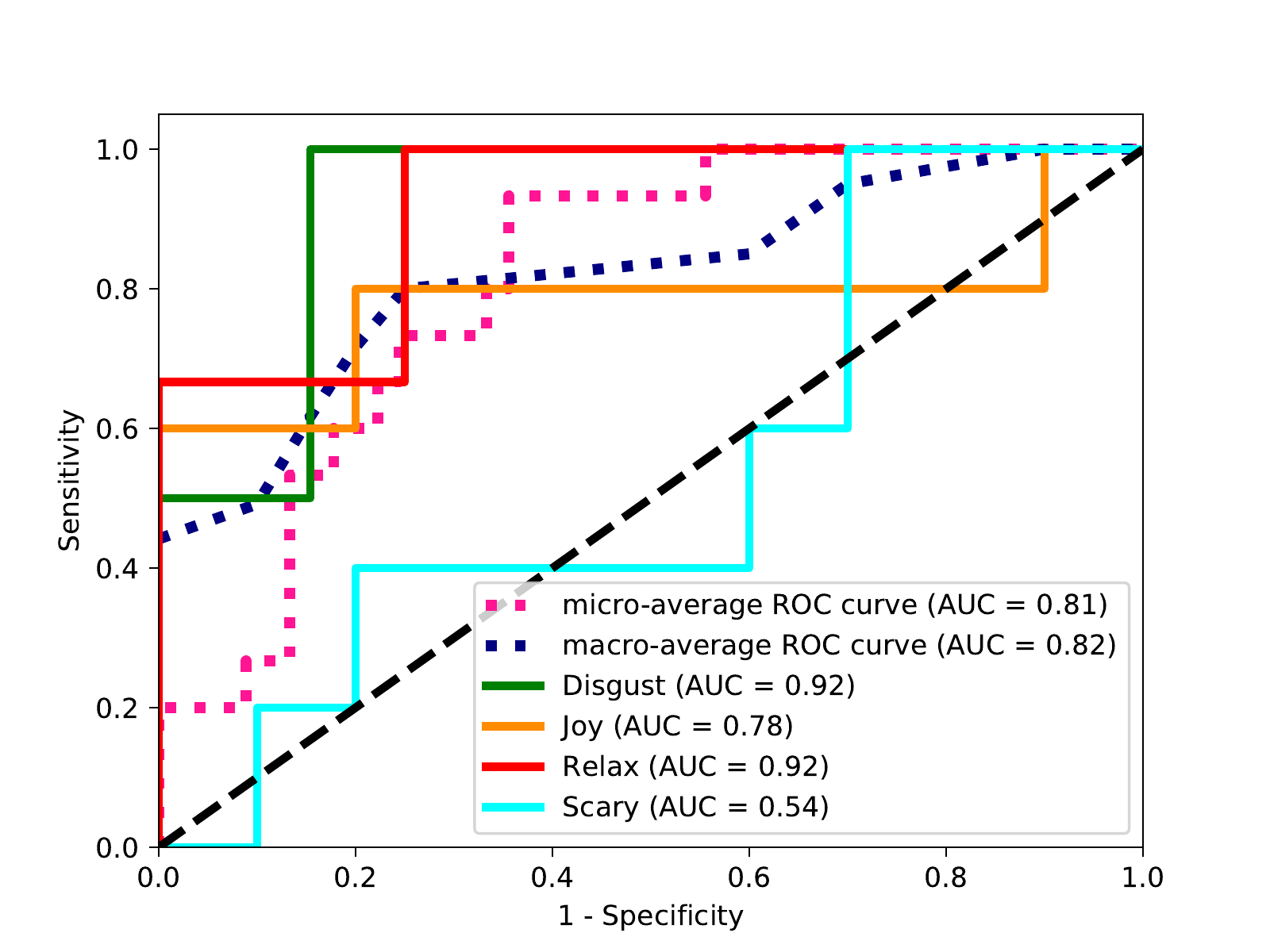} 
		\caption[caption]{ROC-AUC representing the degree of separability between classes. The emotions `Disgust' and `Relax' are highly separable from the rest. Micro-average aggregates the contribution from all classes to compute the average ROC curve. Macro-average computes the ROC metric for each class independently and takes the average, hence treating all classes equally.}
		\label{roc_auc}
	\end{figure*} 
Due to this reason, it is essential to assess the capability of models to distinguish between classes. A receiver operating characteristic (ROC) curve is a probability curve obtained by plotting sensitivity against (1-specificity). Area under the curve (AUC) represents the degree of separability. ROC is defined for a binary classifier system, however, can be extended for a multiclass classification by building a single classifier per class, known as one-vs.-rest or one-against-all strategy. ROC curve and AUC for each class obtained using the SVM model are illustrated in Fig.~\ref{roc_auc}. AUCs indicate that the emotions `Disgust' and `Relax' have a higher degree of separability, complying well with the DL and ML classification results. It should be noted that four video stimuli of respective emotions were displayed to the subjects with minimum delay between the videos and hence it is possible for evoked emotions in the preceding video to persist in the initial part of the following video before it completely vanishes. 
\begin{figure*}[h!]
		\centering 
		\includegraphics[width=16cm]{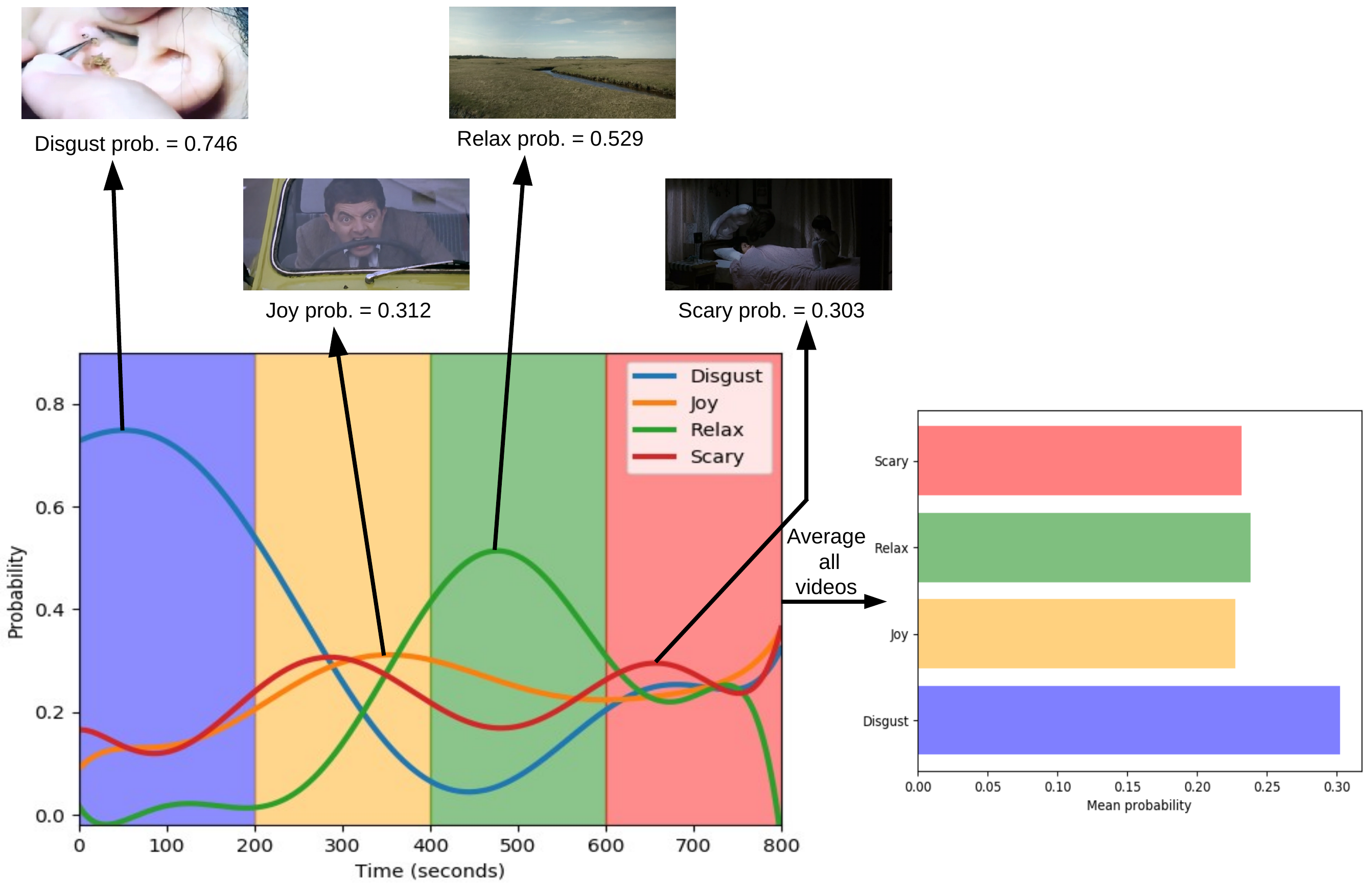} 
		\caption[caption]{Variations of probability of different emotions over time, predicted for a randomly selected subject from the test set. Smooth probability curves are generated by interpolating the discrete probability values.}
		\label{prob_variation}
	\end{figure*}
\\ \noindent
We have used CNN+LSTM model to predict the variations of emotion probabilities across all the videos for a randomly selected subject from the test set. Fig.~\ref{prob_variation} depicts the probability variation of emotions over the time and mean probabilities. 
\newpage
\subsection*{RF vs ECG performance comparison}
\begin{figure}
\begin{subfigure}{1.0\columnwidth}
\centering
\includegraphics[width=1\linewidth]{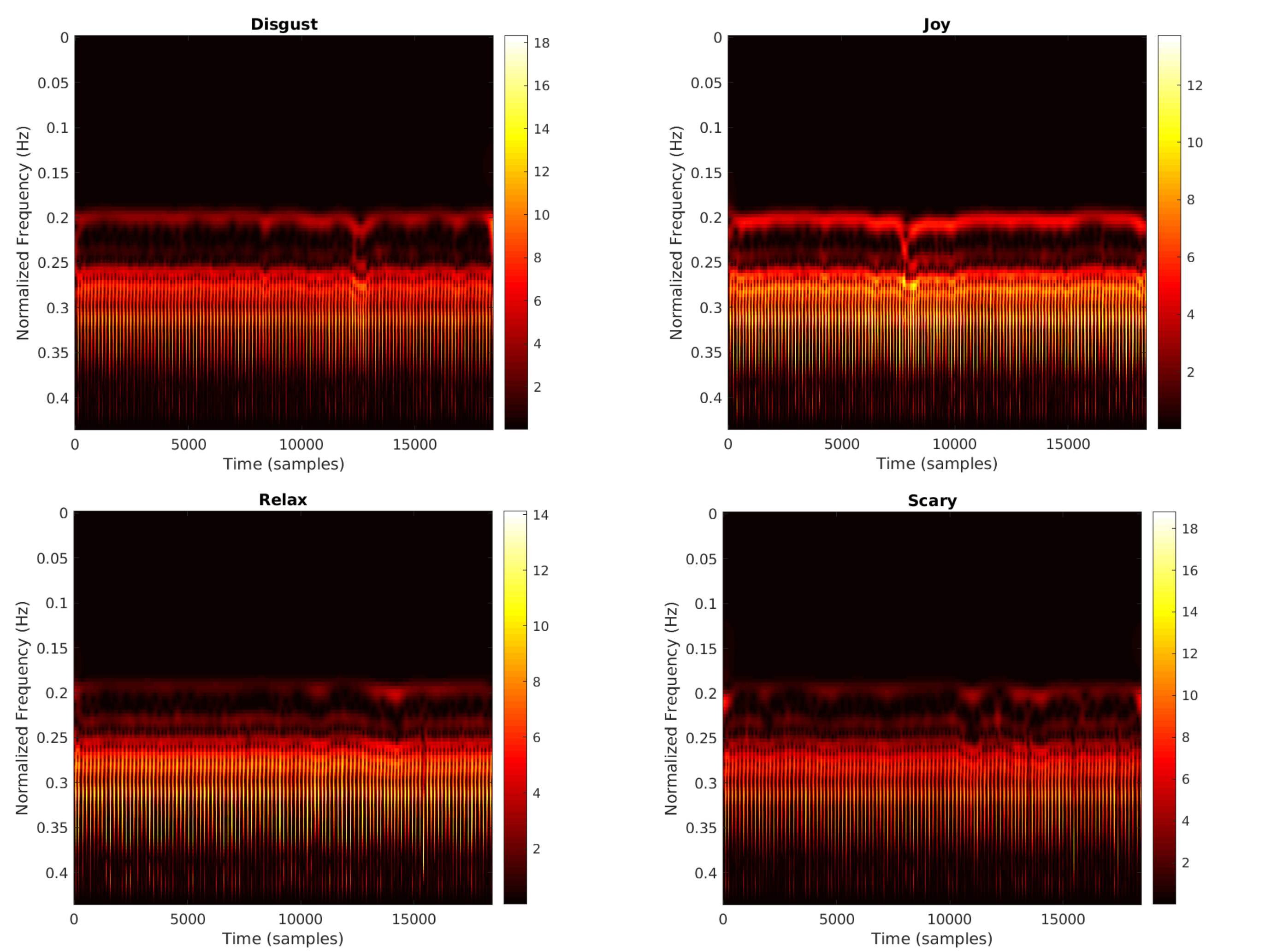}
\caption{CW transformations of different emotions for a randomly selected participant}
\label{ecg_cw}
\end{subfigure}\\
\begin{subfigure}{0.49\columnwidth}
\centering
\includegraphics[width=1\linewidth]{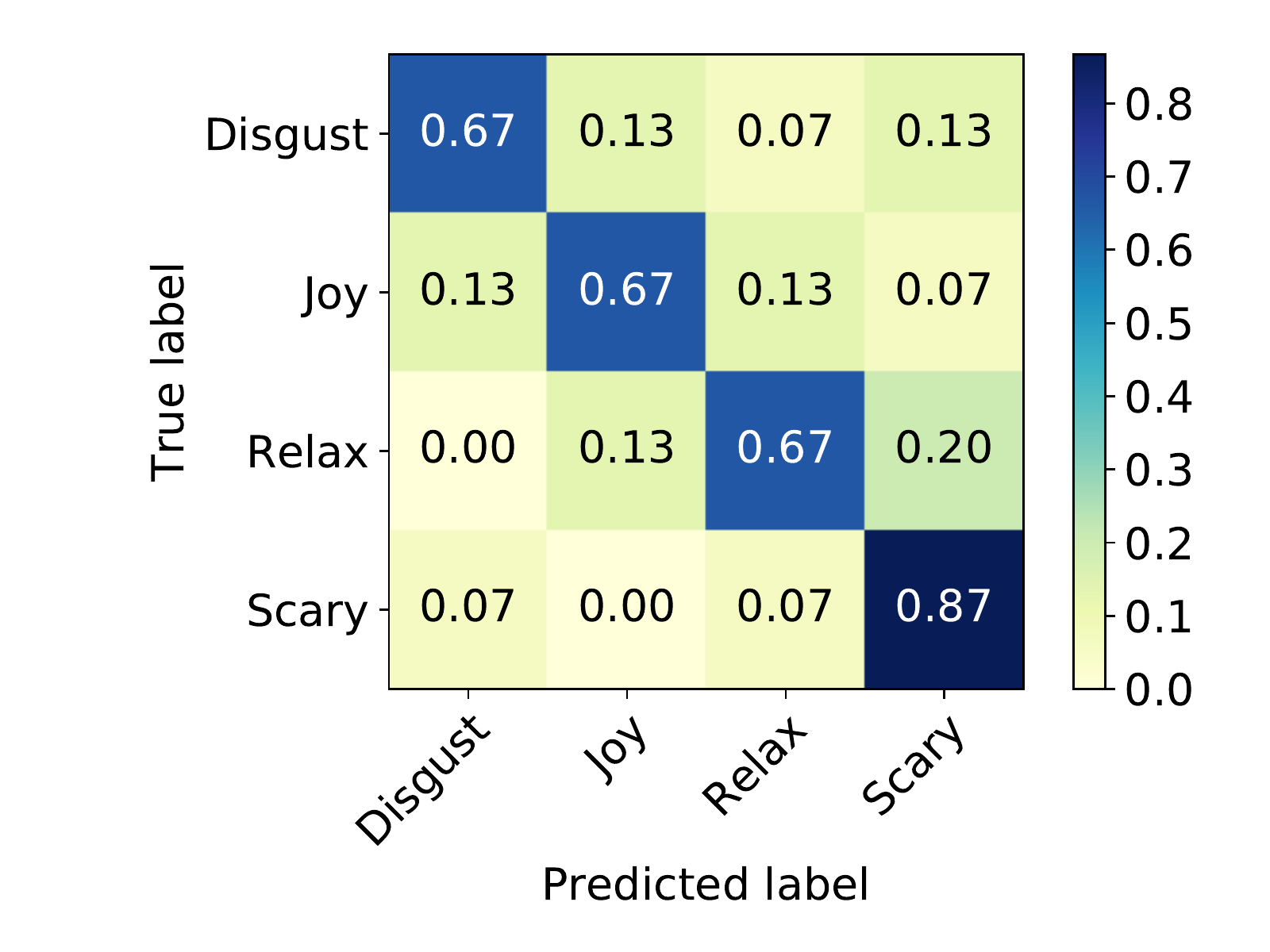}
\caption{CNN}
\label{ecg_dl_conf}
\end{subfigure}
\begin{subfigure}{0.49\columnwidth}
\centering
\includegraphics[width=1\linewidth]{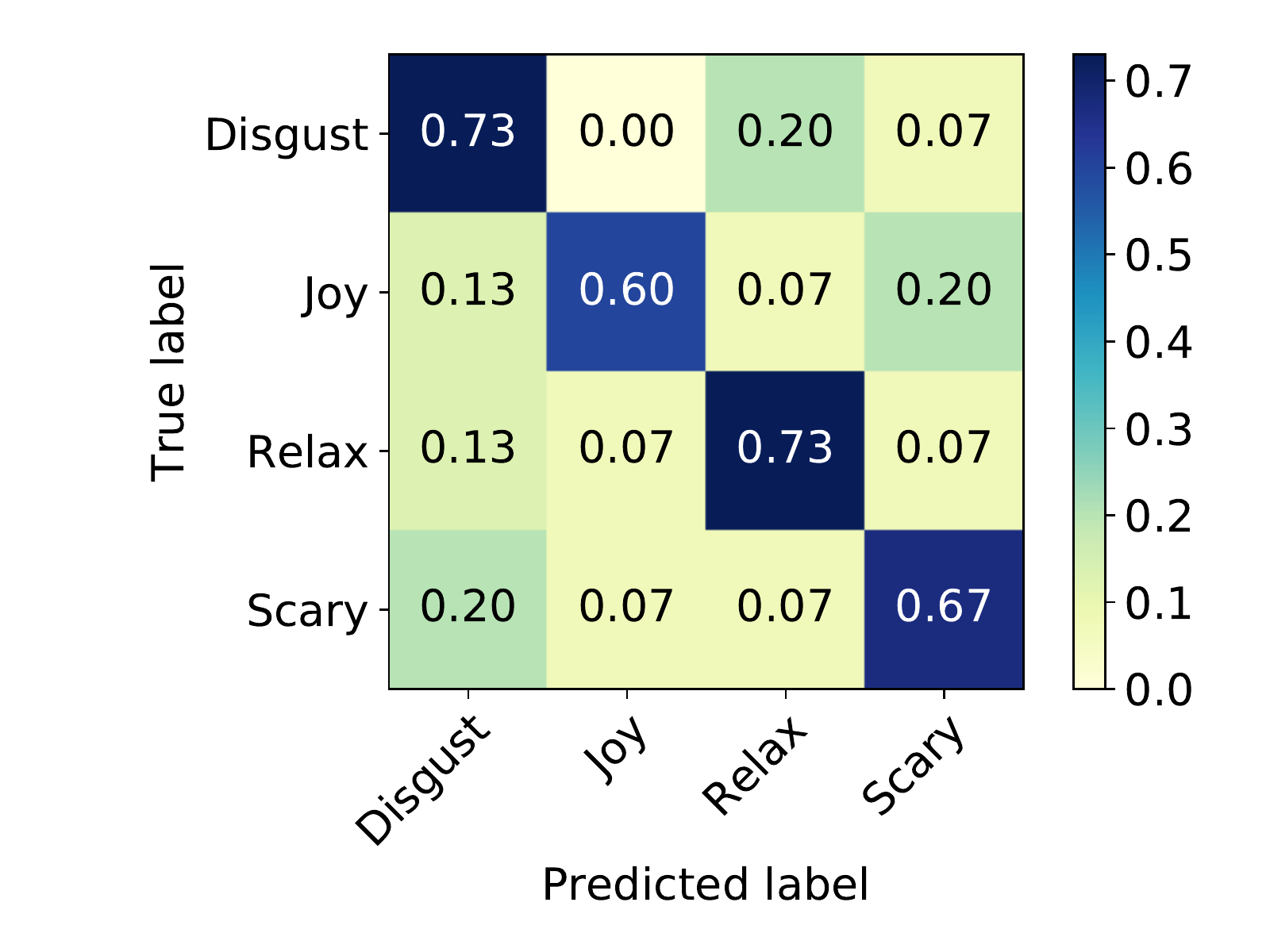}
\caption{SVM}
\label{ecg_svm_cof}
\end{subfigure}
\caption{CW transformed ECG inputs fed to the CNN model and the confusion matrices representing the ECG classification performances of CNN and SVM ML models.}
\label{ecg}
\end{figure}

Human clinical conditions, either physical or mental, cause subtle variations in heart rate that is also reflected in the ECG signal. Therefore, the existing health condition monitoring systems predominantly depend on ECG data for discovering the underlying reasons and categorizing the conditions. In order to make a comparison with RF results, we utilize simultaneously extracted the ECG signal data to train a similar DNN architecture as shown in Fig.~\ref{DL_arch}. Likewise, the Wavelet Transformation is applied to ECG data (Fig.~\ref{ecg_cw}). As observed in the Fig.~\ref{ecg_tsne_plot}, CW transformation alone is not enough to distinguish between emotions. Therefore, we calculated 81 features from the ECG signal, known as inter beat interval (IBI) features and further applied minimum redundancy maximum relevance (mRmR) feature selection \cite{article} method to reduce the dimensions, resulting in 30 features. Since the features do not form a time sequence, we have omitted the LSTM cell from the DL architecture.  The extracted features and the CW transformations are used to train the CNN model for emotion recognition.  To identify the threshold performance, we have trained a SVM model with the extracted 30 features. Table \ref{ecg_perf_metrics} demonstrates the performance metrics in ECG classification. The confusion matrices of CNN and SVM models are shown in Fig.~\ref{ecg_dl_conf} and Fig.~\ref{ecg_svm_cof} respectively. In general, the deep learning classification performances of RF and ECG (Fig.~\ref{ecg}) are high and very similar (both having 71.67\% LOOCV accuracy), indicating that RF signals can describe the underlying emotion of a person as good as an ECG signal with added benefits of being wireless and more practical. Furthermore, we tested the performance of the proposed deep learning architecture on the well established DREAMER ECG database \cite{7887697} for emotion recognition. The model achieved 68.48\% subject independent LOOCV classification accuracy with 0.678, 0.685, 0.680 precision, recall and F1-score values respectively  (supplementary information, section 6). Thus, it is evident that the proposed novel DL architecture can be employed across different databases generated under diverse conditions. 
\begin{table}[t!]
		\caption{{ML vs DL results comparison for ECG classification. The metric `Accuracy' refers to LOOCV accuracy.} }

		\label{ecg_perf_metrics}
		\def\arraystretch{1}
		\ignorespaces 
		\centering 
		\begin{tabulary}{\linewidth}{p{\dimexpr.20\linewidth-2\tabcolsep}p{\dimexpr.20\linewidth-2\tabcolsep}p{\dimexpr.20\linewidth-2\tabcolsep}p{\dimexpr.20\linewidth-2\tabcolsep}p{\dimexpr.20\linewidth-2\tabcolsep}}
			\hline 
			   &  Accuracy (\%) &  Precision &  Recall  & F1-score\\
			\hline 
			CNN &
			71.67 &
			0.720 ($\pm0.03$)&
			0.716 ($\pm0.09$)&
			0.714 ($\pm0.03$)\\
			SVM &
			68.33 &
			0.692 ($\pm0.07$)&
			0.68 ($\pm0.05$)&
			0.681 ($\pm0.02$)\\
			\hline 
		\end{tabulary}\par 
	\end{table}
\section*{Experimental study and data processing}
\subsection*{Ethical Approval}
All experimental study was approved by Queen Mary Ethics of Research Committee of Queen Mary University of London under QMERC2019/25. All research was performed in accordance with guidelines/regulations approved by Ethics of Research Committee. Written informed consent was obtained from the participants involved in the study.
\subsection*{Participants}
The experiment was performed on 15 participants. All participants were English speaking, aged between 22 - 35 years. The participants were briefly explained about the measurement details before the start of the experiment. They were provided with comfortable environment so that they can only focus on watching videos with minimum distractions.
\subsection*{Stimuli: Emotions Evoking Videos}
For inducing emotions in the participants, individual videos were selected that can induce four emotional states (relax, scary, disgust, and joy) in the participants (supplementary informatiion, section 7, Fig. S8). The duration of each video clip was from 3 - 4 minutes. A survey was prepared and provided to the participants, where the emotions can be mapped and graded according to the intensity of emotions felt during the experiment \cite{Nummenmaa646}. Participants were asked to record the intensity of emotions in the survey after watching each video. Self assessment results indicated that videos are capable of inducing a particular emotion in the participant during the experiment (supplementary information, section 5, Fig. S6). However, it is also observed that some participants have experienced multiple emotions while watching a single video. For instance, while watching video corresponding to the happy emotions, participants indicated on the survey that they didn't find the video content happy enough and they remained relax while watching the video. This implies that emotion detection require complex procedure to distinguish emotions of a participant.
\subsection*{Emotion Detection Experiment}
\subsubsection*{Measurement Set-up}
Measurements were performed in the anechoic chamber to reduce any interfering noise emanating from external environment that might alter the emotions of a participant during experiment (supplementary information, section 8, Fig. S11). A pair of Vivaldi type antennas is used to form the radar, operating at 5.8 GHz (supplementary information, section 8, Fig. S10). One antenna is used for RF signal transmission towards the body (Green Signal, Fig.~\ref{Dummy}), while the second antenna was used for receiving RF reflections off the body (Red Signal, Fig.~\ref{Dummy}). A pair of coaxial cables were used to connect both antennas to the programmable vector network analyzer (Rohde \& Schwarz, N5230 C) through coaxial cables. A laptop was used to play videos and the participants were asked to wear headphones so that they can effectively focus on the audio. The distance between the antennas and the participants was 30 cm as illustrated in the measurement set-up (Fig. S2 - supplementary information). 
\subsubsection*{Detection of RF Reflections from the Participants}
The videos were shown one at a time to the participant who was sitting on the chair in-front of the displaying monitor at a distance of approximately 1 meter. The participants were exposed with RF power level of 0 dBm. After the end of each video, the participant was asked to relax before the start of next video.  While each video was playing, RF reflections from the participant's body were detected through the receiving Vivaldi antenna, that was connected the VNA. In our experiment, the phase difference of RF reflections is captured using radar techniques. We have employed the procedure that can calculate the phase difference between the transmitted and RF reflections off the body. For instance, the transmitted signal is given as: \\
$$x\left ( t\right )= \sin \left ( \omega _0t+\varphi _0\right )$$
where $\omega _0$ is the frequency of transmitted signal(operating frequency of 5.8 GHz), whereas $\varphi _0$ is initial phase of the transmitted signal. Distance between the participant and Tx antenna is:
$$d\left ({t}' \right )=d-f\left ( {t}'\right )$$
where $d$ is the static distance between the participant and Tx antenna and $f\left ( {t}'\right )$ corresponds to the movement of participant's body. The received signal can be expressed as:
$$x_{re}\left ( t+\bigtriangleup t\right )=\Gamma \cdot \sin \left ( \omega_{0}t+\omega_{0}\bigtriangleup t+\varphi_{0}\right )$$
\noindent
where $\Delta t=\frac{2d\left ( {t}'\right )}{c}$ is the time duration that the transmitting RF signal takes to reach the participant's body and  $\Gamma =\left |\Gamma _{0} \right |e^{j\varphi _{0}}$ is the reflection coefficient from the participant. By considering the participant's body movement can be regarded as quasi-periodic signals, the expression, $f\left ( {t}'\right )$ can be transformed as $f\left ( {t}'\right )=\sum_{i=0}^{N}A_{i}\sin \left ( \omega_it+\varphi_i  \right )$ . The extended expression of the received signal is given below:
$$
x_{re}\left(t+\bigtriangleup t\right )=\Gamma \cdot \sin\left( \omega_{0}t+\frac{2d}{c}\omega_{0}-\frac{2}{c}\omega_{0}\sum_{i=1}^{N}A_{i}\sin \left (\omega_{i}{t}'+\phi _{i}\right)+\varphi_{0}\right)
$$
The phase difference between transmitted signal and received signal is:
$$
\Phi \left ( {t}'\right )=C_{0}-\frac{2}{c}\omega _{0}\sum_{i=0}^{N}A_{i}\sin \left ( \omega _{i}{t}'+\phi_{i}\right )
$$
\noindent
Where $C_0$ is a constant. The amplitude of $\Phi \left ( {t}'\right )$ is proportional to the frequency of VNA $\omega_0$ and body movements $A_i$. We can infer from above mentioned equation that the variations of phase difference corresponds to the participant's body movement. We have analysed the emotions on the last 120 seconds of each video. This is to make sure that the intensity of emotions will be high by the end of video as compared to the start of every video.
\subsubsection*{Data acquisition using ECG}
The ECG signals have been extensively explored in literature for emotion detection, particularly in the field of affective computing. The emotional states of a person are effectively associated with psychological activities and cognition of humans. In our experiment, we have used an ECG monitor (PC-80B) to extract the heartbeat variations of a participant during experiment. The ECG monitor is convenient to use and has three electrodes that can be mated to the participant's chest conveniently. 
\subsubsection*{Signal Processing Analysis} 
\textit{\textbf{ECG Signals:}} We have employed signal processing techniques on ECG signals to extract the information about heartbeat variation, owning to the  elicited emotions in the participants. Generally, the ECG signals occupy bandwidth in the range of 0.5 - 45 Hz. For this reason, to remove the baseline drift in the ECG signal, re-sampling is applied at the frequency of 154 Hz and a bandpass Butterworth filter is used to perform filtering from 0.5 - 45 Hz. In the next step, we have used Augsburg Biosignal Toolbox (AuBT) of Matlab to extract statistical features from ECG signals for different emotional states (supplementary information, section 9). The extracted features are essential for further classification of emotions. The classification results indicate audio-visual stimulus successfully evoke discrete emotional states and can be recognized in terms of psychological activities.\\
\textit{\textbf{RF Signals:}}
After pre-processing the raw data (supplementary information, section 3, Fig. S5), the next step is to extract feature and transform from processed data. The extracted parameters for ML have been discussed in the previous section, and the transformation based on continuous wavelet transform (CWT) is introduced.\\
For further classifications, we have used continuous wavelet transform (CWT) to modify 1-D RF signals into 2-D scaleogram. In the field of mathematics, CWT is a formal (i.e., non-numerical) tool that provides a complete representation of a signal and provides the capability to continuously alter the scale parameters of wavelets. Based on CWT, the 1-D RF signals can be transformed into 2-D scaleogram that represents an image format. Although a scaleogram is beneficial for in-depth understanding of the dynamic behaviour of body movements, individual body movements of participants while watching videos can also be distinguished individually. The normalized time series and its Fourier transform sequence are extracted as the 1-D features. The 2-D scaleogram that is stored as an image format can be considered as the 2-D features. In the classification section, the combination between 1-D features and 2-D features is used to classify different emotional states of participants.
\section*{Conclusions} 
Emotion detection has emerged as a paramount area of research in neuroscientific studies as well as in many other strands of well-being, especially for mentally ill elderly people that are susceptible to physiological fatigue and undergo interactive therapy for treatment. The emotion detection can augment interactive learning activities that are delivered by the therapists as this approach have the potential in understanding internal body stimuli's that are generated in response to a therapy or treatment, thus ameliorate the effectiveness of therapy. The traditional methods of emotion detection are reliant on the data processing of physiological information, collected by bulky sensors attached to the human body. Moreover, emotions evoked by the audio-visual
stimuli are highly subject dependent and therefore difficult to
classify on a common ground. In this study, we have experimentally demonstrated that a novel deep learning architecture that fuses time-domain wireless raw data with those from the frequency domain can achieve state-of-the-art emotion detection accuracy with good generalizability across different human subjects. The proposed wireless emotion detection system was validated with simultaneously extracted ECG data. Our results indicate that wireless RF signals emitted from via WiFi or Radar for neuroscientific studies of human emotion can be very effective and it offers good detection accuracy in comparison with other alternative approaches. In future work, we will further explore the impact of COVID-19 outbreak on people's emotion by analysing their feelings, emotions and experiences faced during this pandemic. We believe, the framework proposed in the present study is a low-cost, hassle-free solution for carrying emotion related research for healthcare and pandemic situations. 
\section*{Acknowledgements}
This work was supported by IET AF Harvey Research Prize.
\section*{Author contributions statement}
A.N.K contributed in the measurements and wrote the paper (A.A.I and Y.M also contributed in writing). A.A.I. implemented deep learning models and performed results analysis. Y.M. processed data and developed machine learning models. B.L contributed in the measurements. Y.L designed antennas for measurements and Y.H planned and led the research of this
project. All authors reviewed the manuscript.
\bibliography{sample} 

\begin{thebibliography}{10}
\urlstyle{rm}
\expandafter\ifx\csname url\endcsname\relax
  \def\url#1{\texttt{#1}}\fi
\expandafter\ifx\csname urlprefix\endcsname\relax\def\urlprefix{URL }\fi
\expandafter\ifx\csname doiprefix\endcsname\relax\def\doiprefix{DOI: }\fi
\providecommand{\bibinfo}[2]{#2}
\providecommand{\eprint}[2][]{\url{#2}}

\bibitem{10.5555/1204228}
\bibinfo{author}{Hall, P.~S.} \& \bibinfo{author}{Hao, Y.}
\newblock \emph{\bibinfo{title}{Antennas And Propagation for Body-Centric
  Wireless Communications}} (\bibinfo{publisher}{Artech House, Inc.},
  \bibinfo{address}{USA}, \bibinfo{year}{2006}).

\bibitem{6863652}
\bibinfo{author}{{Munoz}, M.~O.}, \bibinfo{author}{{Foster}, R.} \&
  \bibinfo{author}{{Hao}, Y.}
\newblock \bibinfo{journal}{\bibinfo{title}{Exploring physiological parameters
  in dynamic wban channels}}.
\newblock {\emph{\JournalTitle{IEEE Transactions on Antennas and Propagation}}}
  \textbf{\bibinfo{volume}{62}}, \bibinfo{pages}{5268--5281},
  \doiprefix\url{10.1109/TAP.2014.2342751} (\bibinfo{year}{2014}).

\bibitem{8329985}
\bibinfo{author}{{Li}, C.} \emph{et~al.}
\newblock \bibinfo{journal}{\bibinfo{title}{Overview of recent development on
  wireless sensing circuits and systems for healthcare and biomedical
  applications}}.
\newblock {\emph{\JournalTitle{IEEE Journal on Emerging and Selected Topics in
  Circuits and Systems}}} \textbf{\bibinfo{volume}{8}},
  \bibinfo{pages}{165--177}, \doiprefix\url{10.1109/JETCAS.2018.2822684}
  (\bibinfo{year}{2018}).

\bibitem{5738699}
\bibinfo{author}{{Dilmaghani}, R.~S.}, \bibinfo{author}{{Bobarshad}, H.},
  \bibinfo{author}{{Ghavami}, M.}, \bibinfo{author}{{Choobkar}, S.} \&
  \bibinfo{author}{{Wolfe}, C.}
\newblock \bibinfo{journal}{\bibinfo{title}{Wireless sensor networks for
  monitoring physiological signals of multiple patients}}.
\newblock {\emph{\JournalTitle{IEEE Transactions on Biomedical Circuits and
  Systems}}} \textbf{\bibinfo{volume}{5}}, \bibinfo{pages}{347--356},
  \doiprefix\url{10.1109/TBCAS.2011.2114661} (\bibinfo{year}{2011}).

\bibitem{WANG2014406}
\bibinfo{author}{Wang, X.}, \bibinfo{author}{Le, D.}, \bibinfo{author}{Cheng,
  H.} \& \bibinfo{author}{Xie, C.}
\newblock \bibinfo{journal}{\bibinfo{title}{All-ip wireless sensor networks for
  real-time patient monitoring}}.
\newblock {\emph{\JournalTitle{Journal of Biomedical Informatics}}}
  \textbf{\bibinfo{volume}{52}}, \bibinfo{pages}{406 -- 417},
  \doiprefix\url{https://doi.org/10.1016/j.jbi.2014.08.002}
  (\bibinfo{year}{2014}).
\newblock \bibinfo{note}{Special Section: Methods in Clinical Research
  Informatics}.

\bibitem{dias2018wearable}
\bibinfo{author}{Dias, D.} \& \bibinfo{author}{Paulo Silva~Cunha, J.}
\newblock \bibinfo{journal}{\bibinfo{title}{Wearable health devices—vital
  sign monitoring, systems and technologies}}.
\newblock {\emph{\JournalTitle{Sensors}}} \textbf{\bibinfo{volume}{18}},
  \bibinfo{pages}{2414} (\bibinfo{year}{2018}).

\bibitem{doi:10.1002/adma.201301921}
\bibinfo{author}{Jeong, J.-W.} \emph{et~al.}
\newblock \bibinfo{journal}{\bibinfo{title}{Materials and optimized designs for
  human-machine interfaces via epidermal electronics}}.
\newblock {\emph{\JournalTitle{Advanced Materials}}}
  \textbf{\bibinfo{volume}{25}}, \bibinfo{pages}{6839--6846}
  (\bibinfo{year}{2013}).

\bibitem{s101210837}
\bibinfo{author}{Yilmaz, T.}, \bibinfo{author}{Foster, R.} \&
  \bibinfo{author}{Hao, Y.}
\newblock \bibinfo{journal}{\bibinfo{title}{Detecting vital signs with wearable
  wireless sensors}}.
\newblock {\emph{\JournalTitle{Sensors}}} \textbf{\bibinfo{volume}{10}},
  \bibinfo{pages}{10837--10862}, \doiprefix\url{10.3390/s101210837}
  (\bibinfo{year}{2010}).

\bibitem{schwartz2013flexible}
\bibinfo{author}{Schwartz, G.} \emph{et~al.}
\newblock \bibinfo{journal}{\bibinfo{title}{Flexible polymer transistors with
  high pressure sensitivity for application in electronic skin and health
  monitoring}}.
\newblock {\emph{\JournalTitle{Nature communications}}}
  \textbf{\bibinfo{volume}{4}}, \bibinfo{pages}{1859} (\bibinfo{year}{2013}).

\bibitem{webb2013ultrathin}
\bibinfo{author}{Webb, R.~C.} \emph{et~al.}
\newblock \bibinfo{journal}{\bibinfo{title}{Ultrathin conformal devices for
  precise and continuous thermal characterization of human skin}}.
\newblock {\emph{\JournalTitle{Nature materials}}}
  \textbf{\bibinfo{volume}{12}}, \bibinfo{pages}{938} (\bibinfo{year}{2013}).

\bibitem{zhao2016emotion}
\bibinfo{author}{Zhao, M.}, \bibinfo{author}{Adib, F.} \&
  \bibinfo{author}{Katabi, D.}
\newblock \bibinfo{title}{Emotion recognition using wireless signals}.
\newblock In \emph{\bibinfo{booktitle}{Proceedings of the 22nd Annual
  International Conference on Mobile Computing and Networking}},
  \bibinfo{pages}{95--108} (\bibinfo{organization}{ACM}, \bibinfo{year}{2016}).

\bibitem{8105799}
\bibinfo{author}{{Hossain}, M.~S.} \& \bibinfo{author}{{Muhammad}, G.}
\newblock \bibinfo{journal}{\bibinfo{title}{Emotion-aware connected healthcare
  big data towards 5g}}.
\newblock {\emph{\JournalTitle{IEEE Internet of Things Journal}}}
  \textbf{\bibinfo{volume}{5}}, \bibinfo{pages}{2399--2406},
  \doiprefix\url{10.1109/JIOT.2017.2772959} (\bibinfo{year}{2018}).

\bibitem{5738690}
\bibinfo{author}{{Chanel}, G.}, \bibinfo{author}{{Rebetez}, C.},
  \bibinfo{author}{{Bétrancourt}, M.} \& \bibinfo{author}{{Pun}, T.}
\newblock \bibinfo{journal}{\bibinfo{title}{Emotion assessment from
  physiological signals for adaptation of game difficulty}}.
\newblock {\emph{\JournalTitle{IEEE Transactions on Systems, Man, and
  Cybernetics - Part A: Systems and Humans}}} \textbf{\bibinfo{volume}{41}},
  \bibinfo{pages}{1052--1063}, \doiprefix\url{10.1109/TSMCA.2011.2116000}
  (\bibinfo{year}{2011}).

\bibitem{Dolan1191}
\bibinfo{author}{Dolan, R.~J.}
\newblock \bibinfo{journal}{\bibinfo{title}{Emotion, cognition, and behavior}}.
\newblock {\emph{\JournalTitle{Science}}} \textbf{\bibinfo{volume}{298}},
  \bibinfo{pages}{1191--1194}, \doiprefix\url{10.1126/science.1076358}
  (\bibinfo{year}{2002}).

\bibitem{shu2018review}
\bibinfo{author}{Shu, L.} \emph{et~al.}
\newblock \bibinfo{journal}{\bibinfo{title}{A review of emotion recognition
  using physiological signals}}.
\newblock {\emph{\JournalTitle{Sensors}}} \textbf{\bibinfo{volume}{18}},
  \bibinfo{pages}{2074} (\bibinfo{year}{2018}).

\bibitem{rosenkranz2003affective}
\bibinfo{author}{Rosenkranz, M.~A.} \emph{et~al.}
\newblock \bibinfo{journal}{\bibinfo{title}{Affective style and in vivo immune
  response: neurobehavioral mechanisms}}.
\newblock {\emph{\JournalTitle{Proceedings of the National Academy of
  Sciences}}} \textbf{\bibinfo{volume}{100}}, \bibinfo{pages}{11148--11152}
  (\bibinfo{year}{2003}).

\bibitem{verschuere2006psychopathy}
\bibinfo{author}{Verschuere, B.}, \bibinfo{author}{Crombez, G.},
  \bibinfo{author}{Koster, E.} \& \bibinfo{author}{Uzieblo, K.}
\newblock \bibinfo{journal}{\bibinfo{title}{Psychopathy and physiological
  detection of concealed information: A review}}.
\newblock {\emph{\JournalTitle{Psychologica belgica}}}
  \textbf{\bibinfo{volume}{46}} (\bibinfo{year}{2006}).

\bibitem{mandryk2006using}
\bibinfo{author}{Mandryk, R.~L.}, \bibinfo{author}{Inkpen, K.~M.} \&
  \bibinfo{author}{Calvert, T.~W.}
\newblock \bibinfo{journal}{\bibinfo{title}{Using psychophysiological
  techniques to measure user experience with entertainment technologies}}.
\newblock {\emph{\JournalTitle{Behaviour \& information technology}}}
  \textbf{\bibinfo{volume}{25}}, \bibinfo{pages}{141--158}
  (\bibinfo{year}{2006}).

\bibitem{7831367}
\bibinfo{author}{{Mühlbacher-Karrer}, S.} \emph{et~al.}
\newblock \bibinfo{journal}{\bibinfo{title}{A driver state detection
  system—combining a capacitive hand detection sensor with physiological
  sensors}}.
\newblock {\emph{\JournalTitle{IEEE Transactions on Instrumentation and
  Measurement}}} \textbf{\bibinfo{volume}{66}}, \bibinfo{pages}{624--636},
  \doiprefix\url{10.1109/TIM.2016.2640458} (\bibinfo{year}{2017}).

\bibitem{lopez2019towards}
\bibinfo{author}{L{\'o}pez-Hern{\'a}ndez, J.~L.},
  \bibinfo{author}{Gonz{\'a}lez-Carrasco, I.},
  \bibinfo{author}{L{\'o}pez-Cuadrado, J.~L.} \& \bibinfo{author}{Ruiz-Mezcua,
  B.}
\newblock \bibinfo{journal}{\bibinfo{title}{Towards the recognition of the
  emotions of people with visual disabilities through brain--computer
  interfaces}}.
\newblock {\emph{\JournalTitle{Sensors}}} \textbf{\bibinfo{volume}{19}},
  \bibinfo{pages}{2620} (\bibinfo{year}{2019}).

\bibitem{Rudoviceaao6760}
\bibinfo{author}{Rudovic, O.}, \bibinfo{author}{Lee, J.}, \bibinfo{author}{Dai,
  M.}, \bibinfo{author}{Schuller, B.} \& \bibinfo{author}{Picard, R.~W.}
\newblock \bibinfo{journal}{\bibinfo{title}{Personalized machine learning for
  robot perception of affect and engagement in autism therapy}}.
\newblock {\emph{\JournalTitle{Science Robotics}}}
  \textbf{\bibinfo{volume}{3}}, \doiprefix\url{10.1126/scirobotics.aao6760}
  (\bibinfo{year}{2018}).

\bibitem{7536936}
\bibinfo{author}{{Ali}, M.}, \bibinfo{author}{{Mosa}, A.~H.},
  \bibinfo{author}{{Al Machot}, F.} \& \bibinfo{author}{{Kyamakya}, K.}
\newblock \bibinfo{title}{Eeg-based emotion recognition approach for
  e-healthcare applications}.
\newblock In \emph{\bibinfo{booktitle}{2016 Eighth International Conference on
  Ubiquitous and Future Networks (ICUFN)}}, \bibinfo{pages}{946--950},
  \doiprefix\url{10.1109/ICUFN.2016.7536936} (\bibinfo{year}{2016}).

\bibitem{kim2017measuring}
\bibinfo{author}{Kim, J.~J.} \& \bibinfo{author}{Fesenmaier, D.~R.}
\newblock \bibinfo{title}{Measuring human senses and the touristic experience:
  Methods and applications}.
\newblock In \emph{\bibinfo{booktitle}{Analytics in Smart Tourism Design}},
  \bibinfo{pages}{47--63} (\bibinfo{publisher}{Springer},
  \bibinfo{year}{2017}).

\bibitem{kashihara2014brain}
\bibinfo{author}{Kashihara, K.}
\newblock \bibinfo{journal}{\bibinfo{title}{A brain-computer interface for
  potential non-verbal facial communication based on eeg signals related to
  specific emotions}}.
\newblock {\emph{\JournalTitle{Frontiers in neuroscience}}}
  \textbf{\bibinfo{volume}{8}}, \bibinfo{pages}{244} (\bibinfo{year}{2014}).

\bibitem{7727453}
\bibinfo{author}{{Jiahui Pan}}, \bibinfo{author}{{Yuanqing Li}} \&
  \bibinfo{author}{{Jun Wang}}.
\newblock \bibinfo{title}{An eeg-based brain-computer interface for emotion
  recognition}.
\newblock In \emph{\bibinfo{booktitle}{2016 International Joint Conference on
  Neural Networks (IJCNN)}}, \bibinfo{pages}{2063--2067},
  \doiprefix\url{10.1109/IJCNN.2016.7727453} (\bibinfo{year}{2016}).

\bibitem{4468714}
\bibinfo{author}{{Zeng}, Z.}, \bibinfo{author}{{Pantic}, M.},
  \bibinfo{author}{{Roisman}, G.~I.} \& \bibinfo{author}{{Huang}, T.~S.}
\newblock \bibinfo{journal}{\bibinfo{title}{A survey of affect recognition
  methods: Audio, visual, and spontaneous expressions}}.
\newblock {\emph{\JournalTitle{IEEE Transactions on Pattern Analysis and
  Machine Intelligence}}} \textbf{\bibinfo{volume}{31}},
  \bibinfo{pages}{39--58} (\bibinfo{year}{2009}).

\bibitem{Krageleaaw4358}
\bibinfo{author}{Kragel, P.~A.}, \bibinfo{author}{Reddan, M.~C.},
  \bibinfo{author}{LaBar, K.~S.} \& \bibinfo{author}{Wager, T.~D.}
\newblock \bibinfo{journal}{\bibinfo{title}{Emotion schemas are embedded in the
  human visual system}}.
\newblock {\emph{\JournalTitle{Science Advances}}}
  \textbf{\bibinfo{volume}{5}}, \doiprefix\url{10.1126/sciadv.aaw4358}
  (\bibinfo{year}{2019}).

\bibitem{kahou2016emonets}
\bibinfo{author}{Kahou, S.~E.} \emph{et~al.}
\newblock \bibinfo{journal}{\bibinfo{title}{Emonets: Multimodal deep learning
  approaches for emotion recognition in video}}.
\newblock {\emph{\JournalTitle{Journal on Multimodal User Interfaces}}}
  \textbf{\bibinfo{volume}{10}}, \bibinfo{pages}{99--111}
  (\bibinfo{year}{2016}).

\bibitem{TARNOWSKI20171175}
\bibinfo{author}{Tarnowski, P.}, \bibinfo{author}{Kołodziej, M.},
  \bibinfo{author}{Majkowski, A.} \& \bibinfo{author}{Rak, R.~J.}
\newblock \bibinfo{journal}{\bibinfo{title}{Emotion recognition using facial
  expressions}}.
\newblock {\emph{\JournalTitle{Procedia Computer Science}}}
  \textbf{\bibinfo{volume}{108}}, \bibinfo{pages}{1175 -- 1184},
  \doiprefix\url{https://doi.org/10.1016/j.procs.2017.05.025}
  (\bibinfo{year}{2017}).
\newblock \bibinfo{note}{International Conference on Computational Science,
  ICCS 2017, 12-14 June 2017, Zurich, Switzerland}.

\bibitem{ELAYADI2011572}
\bibinfo{author}{Ayadi, M.~E.}, \bibinfo{author}{Kamel, M.~S.} \&
  \bibinfo{author}{Karray, F.}
\newblock \bibinfo{journal}{\bibinfo{title}{Survey on speech emotion
  recognition: Features, classification schemes, and databases}}.
\newblock {\emph{\JournalTitle{Pattern Recognition}}}
  \textbf{\bibinfo{volume}{44}}, \bibinfo{pages}{572 -- 587},
  \doiprefix\url{https://doi.org/10.1016/j.patcog.2010.09.020}
  (\bibinfo{year}{2011}).

\bibitem{8887272}
\bibinfo{author}{{Gu}, Y.} \emph{et~al.}
\newblock \bibinfo{journal}{\bibinfo{title}{Emosense: Computational
  intelligence driven emotion sensing via wireless channel data}}.
\newblock {\emph{\JournalTitle{IEEE Transactions on Emerging Topics in
  Computational Intelligence}}} \bibinfo{pages}{1--11},
  \doiprefix\url{10.1109/TETCI.2019.2902438} (\bibinfo{year}{2019}).

\bibitem{6889383}
\bibinfo{author}{{Ren}, Y.} \& \bibinfo{author}{{Wu}, Y.}
\newblock \bibinfo{title}{Convolutional deep belief networks for feature
  extraction of eeg signal}.
\newblock In \emph{\bibinfo{booktitle}{2014 International Joint Conference on
  Neural Networks (IJCNN)}}, \bibinfo{pages}{2850--2853},
  \doiprefix\url{10.1109/IJCNN.2014.6889383} (\bibinfo{year}{2014}).

\bibitem{kohavi1995study}
\bibinfo{author}{Kohavi, R.} \emph{et~al.}
\newblock \bibinfo{title}{A study of cross-validation and bootstrap for
  accuracy estimation and model selection}.
\newblock In \emph{\bibinfo{booktitle}{Ijcai}}, vol.~\bibinfo{volume}{14},
  \bibinfo{pages}{1137--1145} (\bibinfo{organization}{Montreal, Canada},
  \bibinfo{year}{1995}).

\bibitem{7887697}
\bibinfo{author}{{Katsigiannis}, S.} \& \bibinfo{author}{{Ramzan}, N.}
\newblock \bibinfo{journal}{\bibinfo{title}{Dreamer: A database for emotion
  recognition through eeg and ecg signals from wireless low-cost off-the-shelf
  devices}}.
\newblock {\emph{\JournalTitle{IEEE Journal of Biomedical and Health
  Informatics}}} \textbf{\bibinfo{volume}{22}}, \bibinfo{pages}{98--107},
  \doiprefix\url{10.1109/JBHI.2017.2688239} (\bibinfo{year}{2018}).

\bibitem{6320605}
\bibinfo{author}{{Ghasemzadeh}, H.}, \bibinfo{author}{{Ostadabbas}, S.},
  \bibinfo{author}{{Guenterberg}, E.} \& \bibinfo{author}{{Pantelopoulos}, A.}
\newblock \bibinfo{journal}{\bibinfo{title}{Wireless medical-embedded systems:
  A review of signal-processing techniques for classification}}.
\newblock {\emph{\JournalTitle{IEEE Sensors Journal}}}
  \textbf{\bibinfo{volume}{13}}, \bibinfo{pages}{423--437},
  \doiprefix\url{10.1109/JSEN.2012.2222572} (\bibinfo{year}{2013}).

\bibitem{SABETI2009263}
\bibinfo{author}{Sabeti, M.}, \bibinfo{author}{Katebi, S.} \&
  \bibinfo{author}{Boostani, R.}
\newblock \bibinfo{journal}{\bibinfo{title}{Entropy and complexity measures for
  eeg signal classification of schizophrenic and control participants}}.
\newblock {\emph{\JournalTitle{Artificial Intelligence in Medicine}}}
  \textbf{\bibinfo{volume}{47}}, \bibinfo{pages}{263 -- 274},
  \doiprefix\url{https://doi.org/10.1016/j.artmed.2009.03.003}
  (\bibinfo{year}{2009}).

\bibitem{davidson2003affective}
\bibinfo{author}{Davidson, R.~J.}
\newblock \bibinfo{journal}{\bibinfo{title}{Affective neuroscience and
  psychophysiology: toward a synthesis}}.
\newblock {\emph{\JournalTitle{Psychophysiology}}}
  \textbf{\bibinfo{volume}{40}}, \bibinfo{pages}{655--665}
  (\bibinfo{year}{2003}).

\bibitem{article}
\bibinfo{author}{Ramírez-Gallego, S.} \emph{et~al.}
\newblock \bibinfo{journal}{\bibinfo{title}{Fast-mrmr: Fast minimum redundancy
  maximum relevance algorithm for high-dimensional big data: Fast-mrmr
  algorithm for big data}}.
\newblock {\emph{\JournalTitle{International Journal of Intelligent Systems}}}
  \doiprefix\url{10.1002/int.21833} (\bibinfo{year}{2016}).

\bibitem{Nummenmaa646}
\bibinfo{author}{Nummenmaa, L.}, \bibinfo{author}{Glerean, E.},
  \bibinfo{author}{Hari, R.} \& \bibinfo{author}{Hietanen, J.~K.}
\newblock \bibinfo{journal}{\bibinfo{title}{Bodily maps of emotions}}.
\newblock {\emph{\JournalTitle{Proceedings of the National Academy of
  Sciences}}} \textbf{\bibinfo{volume}{111}}, \bibinfo{pages}{646--651},
  \doiprefix\url{10.1073/pnas.1321664111} (\bibinfo{year}{2014}).

\end{thebibliography}
\end{document}